\documentclass[journal=jacsat,manuscript=article]{achemso}
\usepackage[version=3]{mhchem} 

\usepackage{color}
\usepackage{blindtext}
\usepackage{hyperref}
\usepackage{listings}
\usepackage{xcolor}
\usepackage{enumitem}

\definecolor{codegreen}{rgb}{0,0.6,0}
\definecolor{codegray}{rgb}{0.5,0.5,0.5}
\definecolor{codepurple}{rgb}{0.58,0,0.82}
\definecolor{backcolour}{rgb}{0.95,0.95,0.92}

\lstdefinestyle{bash_colour}{
    backgroundcolor=\color{backcolour},   
    commentstyle=\color{codegreen},
    keywordstyle=\color{magenta},
    numberstyle=\tiny\color{codegray},
    stringstyle=\color{codepurple},
    basicstyle=\ttfamily\footnotesize,
    breakatwhitespace=false,         
    breaklines=true,                 
    captionpos=b,                    
    keepspaces=true,                 
    numbers=left,                    
    numbersep=5pt,                  
    showspaces=false,                
    showstringspaces=false,
    showtabs=false,                  
    tabsize=2
}

\SectionNumbersOn
\setkeys{acs}{articletitle = true}

\author{Idil Ismail}
\email{Idil.Ismail@warwick.ac.uk}
\affiliation[chemistry]{Department of Physics, University of Warwick, Coventry, CV4 7AL, United Kingdom}
\alsoaffiliation[hetysys]{EPSRC Centre for Doctoral Training in Modelling of Heterogeneous Systems, University of Warwick, Coventry, CV4 7AL, United Kingdom}

\author{Shayantan Chaudhuri}
\affiliation[chemistry]{Department of Chemistry, University of Warwick, Coventry, CV4 7AL, United Kingdom}
\alsoaffiliation[dst]{EPSRC Centre for Doctoral Training in Diamond Science and Technology, University of Warwick, Coventry, CV4 7AL, United Kingdom}
\alsoaffiliation[nottingham]{Current address: School of Chemistry, University of Nottingham, Nottingham, NG7 2RD, United Kingdom}

\author{Dylan Morgan}
\affiliation[chemistry]{Department of Physics, University of Warwick, Coventry, CV4 7AL, United Kingdom}
\alsoaffiliation[hetysys]{EPSRC Centre for Doctoral Training in Modelling of Heterogeneous Systems, University of Warwick, Coventry, CV4 7AL, United Kingdom}

\author{Christopher D. Woodgate}
\affiliation[chemistry]{Department of Physics, University of Warwick, Coventry, CV4 7AL, United Kingdom}
\alsoaffiliation[hetysys]{EPSRC Centre for Doctoral Training in Modelling of Heterogeneous Systems, University of Warwick, Coventry, CV4 7AL, United Kingdom}

\author{Ziad Fakhoury}
\affiliation[chemistry]{Department of Physics, University of Warwick, Coventry, CV4 7AL, United Kingdom}
\alsoaffiliation[hetysys]{EPSRC Centre for Doctoral Training in Modelling of Heterogeneous Systems, University of Warwick, Coventry, CV4 7AL, United Kingdom}

\author{James M. Targett}
\affiliation[chemistry]{Department of Physics, University of Warwick, Coventry, CV4 7AL, United Kingdom}
\alsoaffiliation[hetysys]{EPSRC Centre for Doctoral Training in Modelling of Heterogeneous Systems, University of Warwick, Coventry, CV4 7AL, United Kingdom}

\author{Charlie Pilgrim}
\affiliation[mathsys]{EPSRC Mathematics for Real-World Systems Centre for Doctoral Training, University of Warwick, Coventry, CV4 7AL, United Kingdom}

\author{Carlo Maino}
\affiliation[chemistry]{Department of Physics, University of Warwick, Coventry, CV4 7AL, United Kingdom}
\alsoaffiliation[hetysys]{EPSRC Centre for Doctoral Training in Modelling of Heterogeneous Systems, University of Warwick, Coventry, CV4 7AL, United Kingdom}

\title{Eat, Sleep, Code, Repeat: Tips for Early-Career Researchers in Computational Science}

\begin{document}

\begin{abstract}
\singlespace
\noindent 
This article is intended as a guide for new graduate students in the field of computational science. With the increasing influx of students from diverse backgrounds joining the ever-popular field, this short guide aims to help students navigate through the various computational techniques that they are likely to encounter during their studies. These techniques span from Bash scripting and scientific programming to machine learning, among other areas. This paper is divided into ten sections, each introducing a different computational method. To enhance readability, we have adopted a casual and instructive tone, and included code snippets where relevant. Please note that due to the introductory nature of this article, it is not intended to be exhaustive; instead, we direct readers to a list of references to expand their knowledge of the techniques discussed within the paper. It is likely that this article will continue to evolve with time, and as such, we advise readers to seek the latest version. Finally, readers should note this article serves as an extension to our student-led seminar series, with additional resources and videos available at \url{https://computationaltoolkit.github.io/} for reference.

\end{abstract}

\begin{figure}[H]
  \centering
  \includegraphics[width=1.0\textwidth]{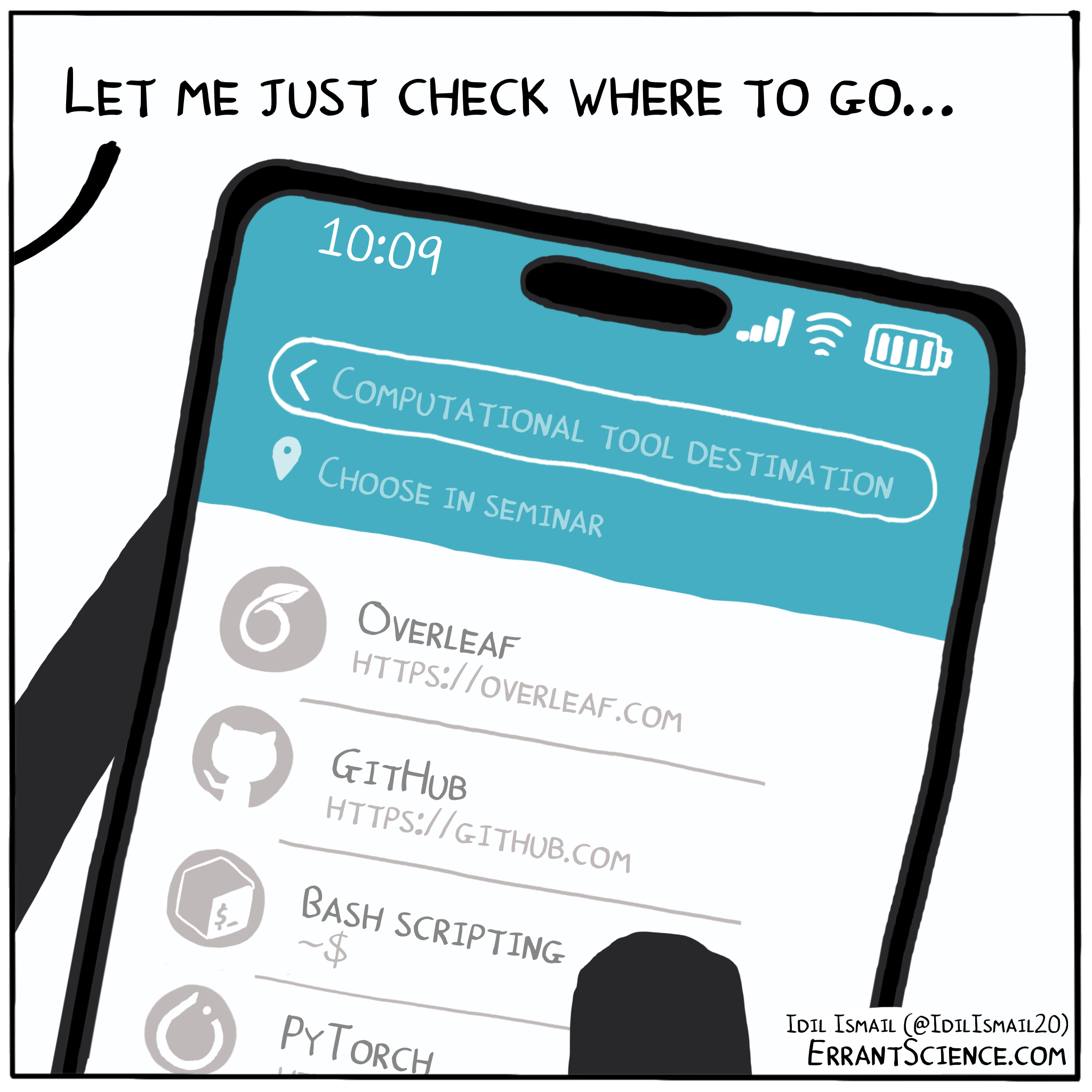}

\end{figure}
\newpage
\tableofcontents
\newpage

\section{Introduction}

Computational science is a highly interdisciplinary field encompassing a wide array of subject areas, including physics, chemistry, engineering, mathematics, life sciences, and more. Our goal in this article is to highlight several uniting themes, skills, and methods used by computational scientists within their respective fields. Serving primarily as an introductory guide for early-career researchers, this article is a collaborative effort, where each section is written by a current graduate student, each bringing forth their unique insights into the topic. While our goal is to ease the learning process and assist researchers in effectively applying the discussed tools and methodologies, we emphasise that readers should treat this guide as informed advice, drawn from the varied experiences and expertise of its contributors, rather than a definitive instructive manual.

\section{Navigating the Zoo of Computational Techniques}\label{sec:zoo}

Computational science, regardless of the precise discipline, means running scientific software on a computer to obtain meaningful results. But what do we mean by `software'? Simply put, scientific software is any program which takes in data and parameters, does something meaningful with it, and returns some new data.

Rather than author a piece in which I try to cover the whole of computational science in one go, I will focus on my specific sub-discipline of computational physics, electronic structure, and hope to draw out some lessons which might be applicable in other areas, too!

\subsection{Where to Start}

On your research journey, you will inevitably have a number of academic mentors, be they your MSc/PhD project supervisor, a more senior student/researcher in your group, or an external collaborator. The intention is that these people are there to \emph{help} you on your research journey, so don't be afraid to ask for help when you need it. Far too often, a student will struggle for days with a problem when another member of their research group would have known exactly what to do, having encountered that same problem before themselves! So, while it is nice to be able to work independently, it is also important to recognise that your time is valuable and that it is worth asking for help with something that you are struggling with if you feel stuck. 

With my PhD supervisor, I arrange a weekly one-on-one meeting in which I discuss what progress I have made and where I should go next, but I am also able to email her or drop by her office if I have a question between meetings. I like to aim to produce a meaningful graph or piece of data for each meeting. Over the course of four years of PhD study (allowing 8 weeks of the year off for holidays or illness) one graph per week works out at over 180 graphs in total, which hopefully makes thesis-writing a little less scary at the end! As a caveat, I will add that, while this arrangement has worked well for me, different research groups work differently, so you should discuss your own schedule with your supervisor.

When you start on a research project, your supervisor or academic mentor will almost certainly give you something from which to start, be it some papers to read, or maybe some test calculations/simulations to run. The idea here is for you to dip your toe into the water working on a problem for which we already know the answer, to make sure you understand what you're aiming for and what pitfalls there might be along the way. It might be that your research project is methods-based, meaning you are going to develop new algorithms, codes, or other tools to solve a problem that existing techniques can't solve or, at least, can't solve efficiently. Your project might have more of a results/ applications focus, meaning you will be applying existing techniques to new problems to see what can be learnt. Or it might be some combination of the two. Regardless, there will be some `toy' problems with which you can get started, and some literature to review, before you can get your teeth stuck into a problem in earnest. Be sure to take these toy examples seriously and keep any code you write/use stored somewhere safely so you can refer back to it later; it will almost certainly come in handy at some point!

At this early stage, it is unlikely that you have had the ability to choose which codes or computational techniques you use, because you will have been working through some well-understood example problems. But this is reasonable; we always have to understand the state of the art before we aim to go further. For example, I work primarily in the areas of alloy theory and magnetism and, in particular, I study the nature of atomic arrangements in these systems and how those atomic arrangements can influence a system's magnetic properties. Some example problems which I would give someone starting in my area would be to compare the energy per atom of an ordered, intermetallic structure with a disordered structure. (A good example of such a system would be the Cu-Au alloy.) There are many suitable codes and techniques, but one could start with comparing the energy difference per atom obtained for a large supercell using a plane-wave DFT code like CASTEP~\cite{CASTEP} with the same energy difference obtained using a code using an effective medium to represent the disorder, such as the coherent potential approximation as implemented in JuKKR~\cite{JuKKR}. I might also encourage someone new to the field to study the effect the magnetic state can have on some systems. For example, for an alloy containing Fe, care should be taken to treat the magnetic state (non-magnetic, paramagnetic, ferromagnetic) appropriately for the system being studied. A good example might be to compare the energy per atom for bcc Fe in each of those three states, and to verify that the ferromagnetic state is the one of lowest energy.

\subsection{Where to Go}

After you have finished your literature review and had a go with some example problems, it will be time to begin working on genuine research problems. Inevitably, this will mean developing your own scripts/code along with using those written by other people. At this stage, it is wise to follow the advice of your supervisor; they have a lot more experience than you do and will have a good idea of how they would make progress on your research problem.

Follow their advice, and keep an eye on what is currently being done in the literature. Often, inspiration can be drawn from reading articles published by others working in your field. Perhaps they have developed a new technique which you think can be applied to your system, or perhaps they are using a different code/technique to you but on the same system. For example, in my own work, we recently performed a study for which one specific aim was to compare the results obtained using a variety of other techniques with that of our own~\cite{woodgate_interplay_2023}. Such comparisons are both necessary and informative; they tell the community what aspect of a problem one code/technique gets `right', and what aspects it gets `wrong'.

\subsection{Going it Alone}

Once you have a feeling for what you are doing, and the current state of the literature for your research area, it is not unwise to begin asking yourself where your research might go next. By the end of your PhD, you may well be more knowledgeable than your supervisor in the particular research sub-area in which you are working!

With this in mind, starting from your current area of research, ask yourself what the outstanding problems are, and then ask yourself how you might go about tackling them. Interdisciplinary and collaboration is key to solving a large number of scientific problems, so make sure to talk to those outside of your area, too, as they may have encountered the same problem in a different setting. Pursue interesting ideas and make progress where you can; very often seemingly small progressions can impact a field once their implications are understood.

\section{Introduction to Bash Scripting}
\begin{figure}[ht]
    \centering
    \includegraphics[width=\textwidth, trim={3.5cm 0cm 3.5cm 0cm}, clip]{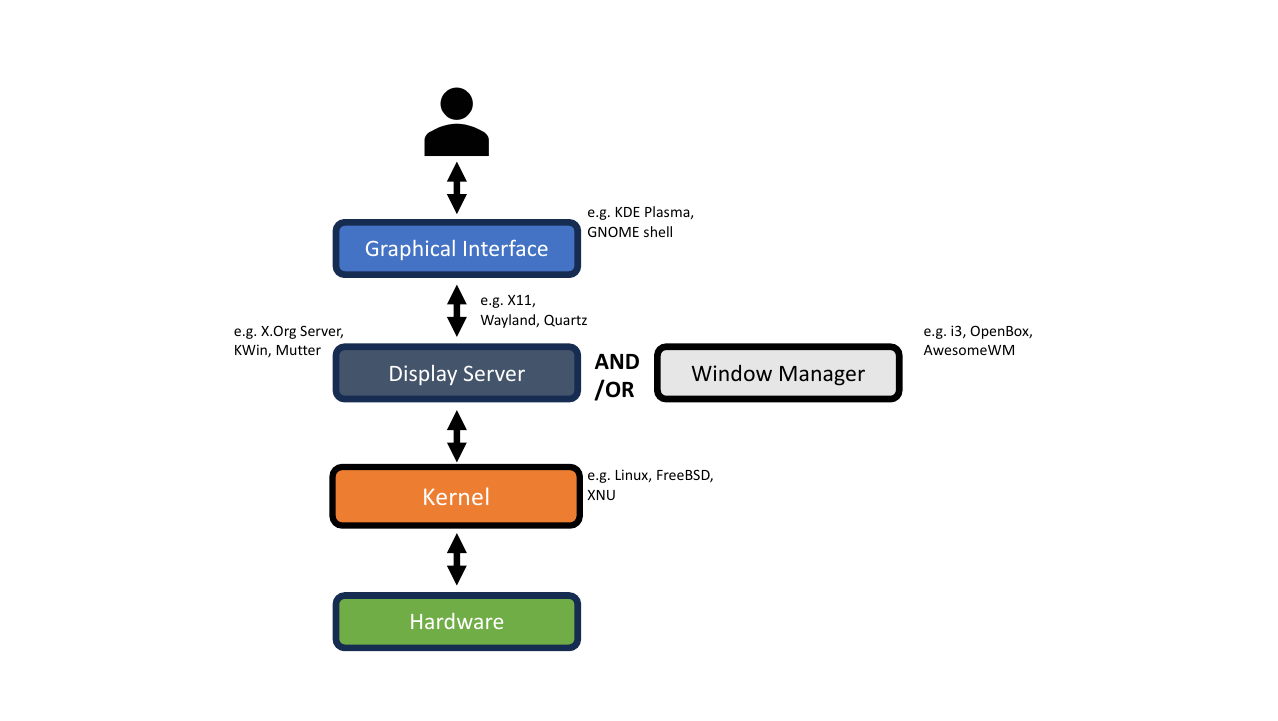}
    \caption{Diagram of the different layers of an operating system and how they interact with each other. Some commonly used examples operating systems can use for each layer are also shown.}
    \label{fig:os_layers}
\end{figure}

It is often necessary in computational science to automate mundane and repetitive tasks. At a somewhat low-level, Bash scripting can be used to achieve this. However, briefly describing an operating system's architecture first will be helpful for understanding this. Beginning at the highest level, the user typically interacts with an application in the graphical interface. The graphical interface is supported by a display server, a window manager, or both, and it interfaces with the display server, for example, on Linux, through X11 or Wayland. Figure \ref{fig:os_layers} shows a diagram of this. The display server is a front end to the kernel, and the kernel is essentially the operating system's source code and what fundamentally differentiates different operating systems from each other. Some well-known examples of kernels include the Linux kernel, FreeBSD, or XNU (for MacOS). 

The command line, accessed through a Terminal emulator (although often incorrectly referred to as just a `Terminal'), provides a direct interface between the user and the kernel and bypasses the graphical interface and display server. A group of languages can be used by a user in a terminal called Shell languages, which use a particular syntax and have scripting capabilities.	

\subsection{What is Bash and Why is it Useful?}
Using a Terminal to control a computer is often the fastest way of accomplishing tasks and, in many cases, especially those that are more complex, the only method. The Bourne Again SHell (Bash) is a Shell scripting language. Shell languages are a group of languages which have a specific syntax to allow the user to use the command line and write scripts. Bash, released in 1989, is arguably the most popular of these and has seen widespread usage across many operating systems, especially in Unix-like OSs. Bash was released to replace the Bourne Shell, which in turn was an improvement to the original sh language. Bash is still the most commonly used Shell today and is almost always the default on Unix-like operating systems (except MacOS as of 2019) and high-performance computing (HPC) systems. HPC clusters do not have a graphical interface, so everything must be done via a Terminal.

\subsection{Useful Features of Bash}

For many individuals beginning a PhD or career in computational science, one of their first tasks is to learn how to navigate their file system. Whilst the goal of this section is not to provide a full tutorial on using Bash (and for that, the reader should refer to Ryans Tutorials\cite{linux_ryan,bash_ryan} on \href{https://ryanstutorials.net/linuxtutorial/}{Linux} and \href{https://ryanstutorials.net/bash-scripting-tutorial/}{Bash} for an excellent introduction), there are several powerful and valuable features I wish to highlight to the reader, as well as use cases and drawbacks in my personal opinion. 

The first is command redirection, with a particular emphasis on piping. Generally, command redirection refers to altering the input and output of commands from the command line. In its most basic form, redirection can be used to write the output of a command to a file or to limit the output to only show error messages, for example. Piping is a slightly more advanced and potent tool for redirection, as it allows the output of one command to be directly used as the input for another using the \verb+ | + symbol. It is commonly employed with another powerful tool, \verb+grep+, a command for searching for text in a larger body of text, like so: \verb+ls -a | grep foo+. This command searches for all files in a directory with the string \verb+foo+ in its name. A more complicated example would be:

\begin{verbatim}
    ls -a | grep foo | xargs cat | awk '{print $4}' | sed 's/pass/fail/g' 
    > foobar.txt
\end{verbatim}

This finds all instances of files in a directory with \verb+foo+ in their name, takes the 4\textsuperscript{th} word in each line, replaces that word with the word `fail' if the 4\textsuperscript{th} word is `pass', and outputs this to a new file called \verb+foobar.txt+. 

Command substitution is similar to redirection, and it involves substituting one bash command in another line of Bash, and the output of the substituted command is used. The following is an example of the use and syntax of this, which sets the variable \verb|fftw_dir| to the location of a file in the \verb|/usr/include/| directory that contains `fftw'. The specific syntax for the command substitution is to place the command in \verb|$()|; however, it is still common to see the now deprecated backtick convention using \verb|``|.

\begin{verbatim}
    fftw_dir=$(find /usr/include/ -name "*fftw*")
\end{verbatim}

At some point, a user will likely come across a case where they need to edit the content of a file. However, how is this possible when there is no GUI available? It would be possible to jointly link several commands directly from the command line; however, this is incredibly unwieldy and arduous. Instead, several applications are often installed on a system that enables editing and viewing files directly in a terminal. The most popular of these today include nano, vim, and emacs. They use different keybindings to navigate through text in files, enabling navigation of the file without a mouse. 

File permission and ownership is another area that often confuses relatively new users of the command line. It may be the case where a user will come across a file they need to read or write and will not have access to it, and Bash provides a particularly non-verbose message of the issue. Listing the directory's contents with \verb+ls -l+ will also output file/directory ownership and permissions information. The second and third columns of the output of \verb+ls -l+ describe its ownership, where the second column is the owner user, and the third column is the group. Ownership can be changed by \verb+chown user_a:user_b path/to/file/or/directory+ where \verb+user_a:user_b+ are the user and group owners, respectively. Permissions are given in the first column of the output of \verb+ls -l+, where there are ten characters for each result. The first is either a `-' or `d', which outlines whether the result is a file or directory, respectively. The next nine characters are organised as three sets of three, each of which refers to the user, the owner, and all users. In this order, each set's character is one of `\texttt{r}', `\texttt{w}', or `\texttt{x}'. These letters indicate whether permissions are set to read, write, or execute the file or directory. If one of the letters has `-' in its place, then the file or directory does not have permission to perform this action. As these are given for the owner, group or all users, it describes which user has which permissions for the file. Ownership can be changed with \verb|chmod ug+w path/to/file/or/directory|, giving the user and group write permissions for the file or directory. An example output using \verb|ls -l| is shown below.

\begin{verbatim}
    total 8
    drwxr-xr-x@ 2 dylanmorgan  staff  64 20 Oct 11:15 bar
    -rw-r--r--@ 1 dylanmorgan  staff   0 20 Oct 11:14 foo
    lrwxr-xr-x@ 1 dylanmorgan  staff  27 20 Oct 11:16 foobar -> ../barfoo
    -rwxr-xr-x@ 1 dylanmorgan  staff  53 20 Oct 11:15 tmp.sh
\end{verbatim}

Finally, documentation is available for every Bash command accessible through the command line. These are called man (manual) pages and are accessed by typing \verb|man|, followed by the command for which the user wishes to check the documentation. When a man page is open, the default vim keybindings (i.e. \texttt{h}, \texttt{j}, \texttt{k}, and \texttt{l}) are used to navigate the page, and `\texttt{q}' to exit and return to the command line.

\subsection{Writing Bash Scripts}

Bash is Turing complete and has features also seen in many other languages, like \verb+if+ statements, \verb+for+ loops, functions, and argument parsing. Beyond just using Bash in a Terminal, Bash is also commonly used to write scripts. Scripts written in Bash have the same functionality and syntax as when using Bash in the Terminal. However, they are helpful when the user wishes to execute many bash commands in succession, schedule tasks, wrappers for other command-line applications, or even for writing command-line programs. The following snippet is an example of a short function taken from a bash script to clean binary files from a compilation of unit tests for a program. 
\newline

\begin{lstlisting}[language=bash]
unit_test_clean () {
    ### Remove compiled binaries from unit testing ###
    bins=("test/test_bin/*.mod" "test/test_bin/*.o" "test/test_bin/*test_spindec")

    # Check for binaries
    if [[ $(ls test/test_bin/* | grep -E 'mod|[.]o|test_spindec') == '' ]]; then
        echo "No binaries found"
        exit 1
    fi

    # Remove globs from $bins if they aren't found
    for glob in ${bins[@]}; do
        if [[ "$glob" == *'*'* ]]; then
            bins=("${bins[@]/$glob}")
        fi
    done

    echo -e "Removing the following files:\n"
    ls test/test_bin/* | grep -E 'mod|[.]o|test_spindec'
    echo

    for file in ${bins[@]}; do
        rm "$file"
    done
    echo "Cleaned successfully"
}
\end{lstlisting}

Bash's widespread use over such a significant period - or at least a long time on a computer time scale - and use in systems administration is helpful for those starting in Bash. Answers to most problems can be found with a quick internet search, and sites such as StackOverflow (\url{https://stackoverflow.com/}) are beneficial for help on specific issues.

Bash is not without its faults, however, and there are many problems that other languages, such as Python or Julia, would be more applicable for. Floating-point variables are not available in standard Bash, so any calculations involving floating-point arithmetic should be avoided, even though they are technically possible to do by piping a string to the \verb|bc| command. It is my opinion that Bash should not be used for any math operations whatsoever. Physically typing Bash scripts can also become a nuisance with the many special characters that must be included. There are also confusing syntax quirks and inconsistencies; for example, sometimes, single brackets should be used over double brackets in \texttt{if} statements. In summary, Bash's strengths lie in argument parsing, easy manipulation of files and directories, and automation of command-line-based commands. 

\section{Introduction to High-Performance Computing}\label{sec:tip13}
HPC~\cite{Dowd98} involves utilising many computing cores in parallel to augment computational performance to tackle complex, large-scale problems that would otherwise be very inefficient and unfeasible. Depending on your work, there is a good chance that, at some point, you will need to use HPC facilities. Such facilities will allow you to run calculations and processes in tandem rather than sequentially. HPC clusters can range in size from local services provided by universities to national and international supercomputers and typically comprise a login node that users can \texttt{ssh} into. This can then be used to submit jobs to a scheduler, which are then executed by compute nodes. Jobs can be run using multiple compute nodes, and each node will typically comprise multiple cores, with some facilities such as the ARCHER2 UK National Supercomputing Service (\url{https://www.archer2.ac.uk/}) providing 128 cores per node. \\

Here are the things, in my opinion, that you should do whenever you get access to a new HPC facility:
\begin{itemize}
    \item \textbf{Set up basic commands:} This includes properly setting up your \texttt{\$HOME/.bashrc} file (or equivalent), which can include useful \texttt{alias} commands. If you like having your own Conda environment, also set this up. It is also good practice to add some \texttt{module load} commands to your \texttt{\$HOME/.bashrc} file for basic modules that might be needed frequently, so these are automatically loaded whenever you are on the login node. This can include programming languages like Python, R, or CMake, often required for software compilation.
    \item \textbf{Familiarise yourself with the HPC:} Most HPC facilities have good online documentation, making it easy to look up anything you need. It is good to familiarise yourself with job resource limits, such as the maximum number of nodes per job/user and the maximum time limit (wall time) per job. It is also important to note that you won't always need the full wall time for every job. Setting the maximum wall time for every job will only increase the queue time for your job before it is executed! It is also worth looking up the different node types that the HPC cluster provides. Many clusters provide different node types, such as high-memory nodes for memory-intensive jobs and developmental nodes for testing, saving you from queuing for the standard compute nodes. Some facilities also include graphics processing unit (GPU) nodes, recommended for ML/ deep learning applications and accelerating software packages~\cite{HuhnCPC20, SpigaIEEE12, VogtJPCA08, WilkinsonJCC13, YanCPC13, GenoveseJCP09}.
    \item \textbf{Software compilation:} When using software to run jobs on a HPC cluster, you must have a parallel binary or library of the software compiled. This will require you to be familiar with the compiler types available on the HPC cluster, e.g. Intel, GNU Compiler Collection (GCC), Cray, and these can typically be loaded as modules. But before compiling anything, always ask around! Do not waste time reinventing the wheel and struggling to compile. Compilations are cluster-specific, so asking your group members for instructions is always worth asking them, as they may have previously compiled the software. Most HPC services have a help desk for compilation and other issues or questions, so don't be afraid to contact them.
    \item \textbf{Submitting jobs:} The most common job scheduler for HPC facilities is Slurm, though some use the Portable Batch System. Ensure you acquire an example job submission script, either from the HPC-specific documentation or from someone who has used the cluster, and submit a test job to ensure everything works. All commands within a job script must be written using the Bash language (or some other Shell variant), so make sure you are aware of simple scheduler commands such as \texttt{srun}/\texttt{qsub}. A good summary of useful Slurm commands can be found at \url{https://slurm.schedmd.com/pdfs/summary.pdf}, though reading through any cluster-specific documentation is still advisable. 
\end{itemize}

\section{Introduction to Machine Learning for Computational Science}\label{sec:chapter6}
In this short section, we aim to introduce the audience to some of the key software and algorithms that, in our opinion, are essential for anyone getting started in machine learning (ML). Given the general scope of this article, we will aim to keep this section brief. Therefore, this will not be an exhaustive list but instead, we will provide references for further reading. The tips and suggestions offered in this section will be rather generic and only really intended for an uninitiated audience. They may not be of much benefit to more seasoned ML practitioners. 

\subsection{What is Machine Learning?}

\begin{figure}[htbp]
  \centering
  \includegraphics[width=0.7\textwidth]{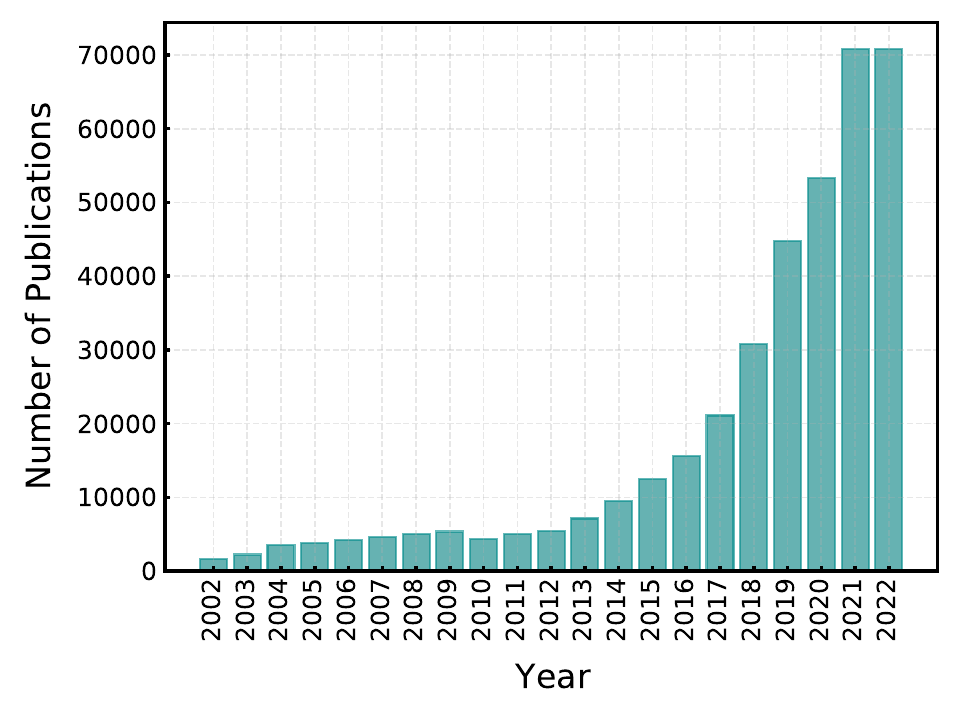}
  \caption{Number of papers published in the field of machine learning (ML) over the past two decades (Web of Science (2023). Retrieved from https://www.webofscience.com)}. 
  \label{fig:barplot}
\end{figure}

With the ever-increasing availability of larger datasets and computing power, a growing number of researchers are embracing ML as an integral part of their computational toolkit, a trend that is further demonstrated by recent reports of over 70,000 ML-related papers being published in 2022 alone (Figure \ref{fig:barplot}). This is not least because of its success in offering significant improvements in speed over more computationally demanding first-principles calculations, helping to further streamline and accelerate the discovery pipeline.

At this stage, it may be beneficial to explain what ML is and to give a short overview of its main principles. Simply put, ML is a sub-field of artificial intelligence concerned with modelling data using algorithms capable of learning and extracting meaningful patterns and relationships between sets of input variables. In each case, the computer learns from previous examples and optimizes its performance on a given task accordingly. A key feature of ML is its ability to make predictions on new data without being explicitly programmed. Broadly speaking, ML can be further subdivided into three main categories; supervised learning, unsupervised learning, and reinforcement learning.\cite{Bishop2006Jan,Murphy2012Sep} Of the three, supervised learning models are the most popular and relevant type of ML in the natural sciences, and will therefore be the focus of the next few sections. 

A key distinguishing feature of supervised ML is that the model is trained on a labeled dataset, where there is a direct and explicit mapping of the input (predictor) and output  (response) variables.\cite{Bishop2006Jan} The goal here is for the model to learn a particular relationship that exists between those two variables. These ML models can be used in either regression or classification tasks, where the former requires continuous data as input and the latter requires a model to be trained on categorical labelled data. Examples of supervised learning models include linear regression, decision trees, and artificial neural networks (ANNs). As the name suggests, linear regression algorithms are used to model linear relationships between variables. However, in cases where the data is non-linear, one might opt for alternative ML models, like decision trees or ANNs, which are capable of modelling complex non-linear relationships between variables.

\subsection{How to Get Started}

There is a general and accepted approach to training and deploying a ML model, which can be summarized in the following steps.

\textbf{Data Acquisition.} In the first instance you will need a large enough dataset generated using a consistent methodology, whether that be experimental or simulated. Note that the reliability of your ML model will largely depend on the quality of your data, so it is important to evaluate and perhaps perform some basic exploratory analysis on your dataset before you get started. For more tips and suggestions on how to select a suitable dataset see \emph{Lones 2021}.\cite{lones_how_2021} 

\textbf{Feature Selection.} When developing any predictive model, it is important to consider the functional relationship between the input feature and the target value you are trying to predict. The procedure by which this is modeled is known as feature selection. There are generally two key points to keep in mind when deciding on which features you want to encode into you ML model.
\begin{itemize}
\item Though this may sound obvious, your chosen feature must be correlated to the target property and should be relevant to the desired prediction.
\item Beware of choosing redundant features. Often, it is tempting to pack your model with as many features as possible, but you should restrict your chosen features to only those that improve the predictive capabilities of the model. Features that do not contribute to the accuracy of the model, often do nothing more other than increase the overall computational cost, but could also contribute to worsening the overall predictive performance by learning redundant information.
\end{itemize}

\textbf{Featurization.} 
Before we can progress with training a ML model, we must first consider how we will represent and ultimately transform our data into a format that a computer can interpret, e.g., a vector representation. Effectively, we need a unique and non-trivial descriptor that can capture all the relevant features of the input data. The requirements for your descriptor will depend on the nature and complexity of your data-type, however, generally, for molecular descriptors in chemistry and physics, one might consider the following key requirements: 
\begin{enumerate}[label=(\roman*)]
    \item \textbf{Unique}: such that structurally different molecules are mapped to distinct vector representations. ML models cannot distinguish between identical features ascribed to different molecules, and will likely give poor or inconsistent predictions in such cases.
    \item \textbf{General}: where the descriptor should be applicable to all atomistic systems. The representations should remain faithful irrespective of whether they are used to encode isotopes, charged, or neutral molecules.
    \item \textbf{Efficient}: with respect to the computational cost associated with training a ML model using geometric descriptors. Given that ML is primarily used to speed up traditional numerical calculations, it is imperative that the ML predictor be orders of magnitude faster than traditional simulations.
\end{enumerate}

Moreover, additional descriptor requirements may involve properties such as rotational, translational, and permutational invariance, to ensure faithfulness of the ascribed feature.\\

\textbf{Model Selection.} The next step involves selecting a suitable ML algorithm that can map given input features onto each other. The model you end up choosing will depend on a number of factors, including but not limited to:

\begin{enumerate}[label=(\roman*)]
    \item Whether your task requires a clustering, classification, or a regression algorithm.
    \item The complexity of your data: Linear regression models will most likely be unsuitable to modelling high dimensional and complex data, which will instead benefit more from the likes of ANNs.
    \item Whether your data is labelled or unlabeled will determine whether you will require a supervised learning or unsupervised learning model.
    \item The size of your dataset: Certain models are better suited for large datasets, such as support vector machines (SVM), random forests (RF), and deep learning models like convolutional neural networks (CNN). In contrast, Gaussian process regression is often restricted to smaller datasets due to its heavy computational costs.\cite{Belyaev2014Mar,Liu2018Jul}
\end{enumerate}

\textbf{Data Splitting.} To ensure that you've built a reliable, and crucially, testable model, you will want to set aside a portion of your dataset for testing, typically somewhere around $~20\%$. The data points selected for testing should be equally distributed, and preferably randomly selected. There are a number of techniques that can be used to achieve this, e.g., K-fold cross validation.\cite{sklearn_api} You should now be able to train your ML model using the remaining $~80\%$ of your dataset. 

\textbf{Hyperparameter Optimization.} Depending on your ML algorithm of choice, you can optimize your model hyperparameters to reduce the overall loss function. There are a myriad of existing software libraries that contain ready made functions you can use to accomplish this. Examples include brute-force algorithms such as grid-search methods (GridSearchCV) offered in Scikit-learn\cite{Scikit-learn} or alternatively, Bayesian optimization methods.

\textbf{Model Training \& Evaluation.} Once your model parameters are refined, you can start training your model. To determine the success and accuracy of your model, you can then test it on the previously saved dataset and evaluate the accuracy of the predictions made using an error metric (loss function) of your choice; popular ones include the likes of the mean absolute error (MAE) and root mean squared error (RMSE). For regression tasks you may also take note of the coefficient of determination (R$^2$). 

\textbf{Model Deployment.} Finally, once you're confident and happy with your testing error metrics, you can deploy your ML model on new and unseen data. This is the ultimate test of the accuracy of your newly trained ML model. Of course it would help to benchmark your results against experimentally/theoretically verified findings for further confirmation.

\subsubsection{Essential Preparations Before Training Your First Machine Learning Model}

Now that you've acquainted yourself with the general workflow of a typical ML task, you might be wondering how these steps can be applied in practice. However, as with learning any new skill, it is essential to have certain prerequisite knowledge. This is important not only to develop your ML model, but also to optimize and interpret its results. Listed below are three of what we deem are the most important prerequisites you should have before venturing to train your first ML model.

\begin{itemize}
    \item \textbf{Mathematics:} First, it is most helpful to have a solid background in at least the basics of linear algebra and calculus. This will enable you to understand the algorithms you are working with, evaluate your dataset, and of course, test your model. 

    \item \textbf{Domain Knowledge:} Similarly, having prior knowledge of the domain on which you intend to apply your ML model is arguably just as important as understanding the workings of your model, as this could highlight whether or not ML may even be suitable for the problem you are tackling.

    \item \textbf{Programming:} Finally, you must have a good grasp of at least one suitable programming language, and preferably Python. Other semi-popular programming languages for ML include Julia, R, and Matlab. However, a general familiarity with data structures and algorithms would also be helpful.
\end{itemize}

\subsubsection{Exploring Popular Machine Learning Packages}
Given the current excitement and widespread interest in ML, we are fortunate to have access to a variety of open-source software packages that can be used to train and deploy a ML model. Due to the popularity of Python in the ML community; which is due in large part to the language's ease and flexibility, many of these software packages are also Python-based. Examples include Scikit-learn,\cite{Scikit-learn} TensorFlow,\cite{TensorFlow} and PyTorch\cite{PyTorch} (\textit{though the core implementation of the last two is technically written in C++}). All three of these packages are well documented, each with their own set of strengths and weaknesses. The package you choose to start you ML journey with will largely depend on the particular requirements of your project. The latter two examples; TensorFlow and PyTorch, are more suited to deep learning tasks, while Scikit-learn offers many standard out-of-the-box supervised and unsupervised ML algorithms, while also complementing other popular libraries including Matplotlib~\cite{Matplotlib}, NumPy~\cite{NumPy}, and SciPy~\cite{SciPy}.\\

\subsection{Should I \textit{really} Be Using Machine Learning for This Problem?}

With all the excitement and hype surrounding ML, it's easy to get caught up in the buzz. However, as computational scientists, we must tread carefully and first consider whether our research problem is worthy of employing ML techniques at all; as the overall accuracy and generalisability of your results rest on this decision. To help you assess the suitability of ML for your research question, here's a checklist of statements to consider before jumping into the use (and abuse) of ML.

\begin{enumerate}

    \item \textbf{Do I have enough quality data to train my ML model?} \\
    ML models are typically data-hungry. Therefore, the first point of inquiry should be whether there is sufficient and relevant data available to train your ML model. Here are some additional points to consider regarding your data requirements:
    \begin{itemize}
        \item Do I need to curate my own dataset, and is it within the computational budget to be considered practical?
        \item Is the data consistent and of high enough quality to ensure reliability?
        \item Is the data already in a format ready to use, or will I need to spend additional time pre-processing the data i.e., cleaning and feature engineering? Again, is it expense practical?
    \end{itemize}

    \item \textbf{Will using ML significantly speed up your calculations?} \\
    This question requires one to first assess the general speed of their calculations using traditional (\emph{possibly first-principles}) simulations. ML is not a silver bullet, and the processes of training and deploying an ML model can take some time itself. Therefore, one must consider factors that may help them decide and evaluate the potential of using ML to accelerate their calculations, and these include:
    \begin{itemize}
        \item Is the time-complexity of your problem high enough to benefit from the computational speed offered by ML? 
        \item Does your research question require the identification of patterns and trends in large datasets?
        \item Does your research question generate clear and measurable outputs?
    \end{itemize}

      \item \textbf{Is my research problem too complex to be modelled using ML?} \\
      To answer this question, first, you need to consider the possible reasons why you believe that traditional numerical simulations or statistical methods cannot adequately address your research aims. These may include questions such as:
    \begin{itemize}
        \item How much domain knowledge do you require to train your ML model? To answer this, you will need to determine whether your investigation requires a direct or inverse approach. The latter likely requires more domain knowledge than the former where the scientific or physical principles are typically already known. An inverse problem research approach involves determining causal factors on the basis of observed simulation output, and therefore inherently requires expert knowledge to explain the physics or chemistry behind the observations, thus making it less suitable for ML investigation.
        \item How complex or intricate are the features in your study? If your problem involves a high-dimensional feature space that is too complex to model via other methods, then its likely that ML may not perform well either; as its known to perform better with fewer and more relevant features. Again, ML is not a cure-all solution.
    \end{itemize}
\end{enumerate}
Finally, really ask yourself if you have trialed and exhausted all other possible statistical and simulation methodologies before turning to ML-based approaches, as these will likely offer far better and more reliable solutions to your problem.

\section{Popular Tools For Scientific Programming in Python}

\subsection{Do Not Reinvent the Wheel}

The computational sciences can be applied to a wide range of disciplines, where many of these fields share common tools and techniques, for example, visualization software, as we all need to plot graphs! Others including ML, as mentioned previously, has been applied to a whole range of disciplines and problems.

Moreover, note, that it is more than likely that you will find an existing piece of software for your specific problem case, so `don't reinvent the wheel' (\emph{if you don't have to}). In fact, it can be quite detrimental at times to rewrite tools that have already been written from an efficiency standpoint. Most of these tools have had teams of people create optimized, shared libraries that will almost definitely be better than a quick implementation you or I can come up with. It is therefore quite vital that before you embark on any code sprints, to check what tools you have at your disposal and how easy or straightforward it would be to incorporate them in your project.

\subsection{Why Python?}

Python is a high level, general purpose, programming interpretative language. This means it is more understandable and flexible. It boasts readability and interoperability with low-level compiled languages (mainly C/C++). It's highly popular amongst different software engineering areas outside of computational science, such as web development, game development, and data analysis.

The popularity of Python within the scientific computing community mainly stems from the fact there is an abundance of easy-to-use, well-documented, and optimized software packages written exclusively in Python. The availability of these packages, alongside interoperability and quick development, makes Python an easy choice for most computational scientists. Therefore, this guide will primarily focus on suggesting Python packages that prove to be useful for computational science tasks. Other high level options do exist and are used by many research groups, including, Julia and R. However, Python remains the clear `go-to' language in this field.

\subsection{Popular General Scientific Python Packages}

\subsubsection{NumPy/SciPy}

One of the trade-offs for Python being such a high-level interpretative language, is its slower line-by-line interpretation. This is especially problematic when it comes to dealing with large matrices or multi-dimensional arrays which may require nested loops.  However, there are Python libraries designed to efficiently address these issues, namely, Numerical Python, or NumPy for short.\cite{NumPy} At the core of the NumPy package lies the array object, which efficiently stores and manipulates these multi-dimensional arrays. It provides a wide range of mathematical functions and operators optimized for numerical operations on arrays, enabling fast and memory-efficient computation through the use of compiled C/Fortran code. Its intuitive syntax and extensive documentation make it easy to use, even for those new to scientific computing. In addition, SciPy\cite{SciPy}, or the Scientific Python Library, builds upon NumPy and provides a vast array of advanced algorithms and functions for scientific computing tasks. It covers areas such as optimization, interpolation, signal and image processing, statistics, and more. SciPy's comprehensive collection of sub-modules includes \texttt{scipy.optimize} for optimization problems, \texttt{scipy.interpolate} for interpolation and smoothing techniques, \texttt{scipy.signal} for signal processing, and \texttt{scipy.stats} for statistical analysis. These functionalities make SciPy a versatile and indispensable tool for solving complex scientific and mathematical problems efficiently. It cannot be  overstated how crucial these two libraries are to the field of computational science. The ease of incorporating these advanced and efficient algorithms into a workflow allows you to focus on the bigger picture and avoid spending time dealing with nasty bugs on a sub-par implementation of these tools. These two libraries are a must have for any computational scientist.

\subsubsection{Matplotlib and Seaborn}

Matplotlib~\cite{Matplotlib} is an essential tool for computational scientists as it provides a powerful and versatile framework for creating high-quality plots and figures. It's flexibility, integration with other scientific libraries, and extensive customization options makes it invaluable for tasks such as data exploration, analysis, and communication in computational science research. 

While Matplotlib is considered quite a low level visualization package in Python, it is highly versatile and allows users to easily customize their plots, it often requires quite a bit of effort to make presentable or publication-quality figures, even for standard plots. 

Similarly, Seaborn~\cite{Seaborn} is a package built on top of Matplotlib that enables quick high quality figure generation for several standard plots. Also, since it's built on top of Matplotlib, many of the customizations you can make through Matplotlib are accessible with Seaborn based plots too.

\subsubsection{Pandas}

Pandas\cite{Pandas_Zenodo} is a Python package that I like to describe as the Python extension of Excel, except it is much quicker, more versatile, and more functional than Excel (and in my opinion, easier to use!). It's a crucial library for data manipulation and analysis. It provides efficient data structures and analysis tools that are essential for processing and exploring large datasets; something you will undoubtedly encounter during your PhD. Pandas also allows you to read data from a large host of formats and stores it in a DataFrame object, storing your data in a tabular format, not very dissimilar for a Microsoft Excel spreadsheet. From this, you have a list of tools for cleaning, reprocessing, querying, quick statistical analysis and visualisation.

Additionally, Pandas makes heavy use of NumPy and is seamlessly integrated with it. This allows us to quickly store the results of complex numerical operations in simple DataFrame objects for quick analysis, providing a smooth and streamlined workflow.

\subsubsection{Jupyter Notebook}

Python's line by line interpretation allows for an interactive and iterative workflow. Being able to run just a few commands, examining their outputs, maybe making a few more adjustments before continuing with another set of commands. This can be quite useful for early exploratory experiments, as we discussed earlier in Section~\ref{sec:zoo}.

Jupyter Notebooks~\cite{Jupyter} are actually a web application, that can be installed like any other Python package. It's set up on a local server that can run interactive blocks of code, such as Python but also Julia and R, and are output straight to the same page. It takes advantage of the iterative interactive nature of Python (again, Julia and R as well) code, to be able to build annotated iterative workflows to be shared with others.

Jupyter Notebooks are valued highly amongst computational scientists, who are constantly trying to share and reproduce scientific work. It is not uncommon to see many presentations and tutorials using Jupyter Notebooks in this area. In some cases, Notebooks have been submitted along published work as supporting material, like the famous LIGO detection of gravitational waves experiment!\cite{Abbott2020Feb}

\subsubsection{Scikit-Learn}

Scikit-learn~ \cite{Scikit-learn} is a popular open-source ML library in Python. It is also built on top of NumPy and SciPy and provides a wide range of algorithms and tools for ML, statistical modeling, and data analysis. Scikit-learn offers popular tools for building classical well established ML models for classification, regression, clustering, and dimensionality reduction. These are all integral tasks for working in predictive modelling and as such, Scikit-learn will probably be your first point of entry when working on any predictive modelling project.

\subsubsection{PyTorch/TensorFlow/Jax}

PyTorch\cite{PyTorch}, TensorFlow,\cite{TensorFlow} and Jax\cite{Jax} are regarded as deep learning packages, although they can be used more generally. Deep learning is a subset of ML that utilises large neural networks, sometimes with billions of parameters to preform predictive tasks. The computational requirements for optimising such models are very different to classical ML methods that can be found in Scikit-learn. Typically, it requires running on specialized hardware like graphical processing units (GPUs) or massive computer clusters. The optimization methods also tend to be numerical iterative algorithms rather than analytic solutions that can found in most classical ML methods. These two computational requirements inspired the development of these deep learning packages. They consist of merging an Autograd framework~\cite{Autograd} with a framework for operating on multi-dimensional arrays on GPUs and accelerated hardware. These requirements mean that PyTorch/TensorFlow/Jax have their own array object, separate to the NumPy array, that have the Autograd capabilities and hardware accelerated compatibility. This allows for efficient implementation of these large scale neural networks using Python. Additionally, these libraries typically offer pre-built neural network architectures. In truth, you can preform any sort of computation using these libraries if the acceleration and Autograd capabilities are useful.

\section{Markup Languages for Scientific Writing}
\label{Section：Markup}

Perhaps you've seen your colleagues produce stunning web pages, documents, or Jupyter notebooks, leaving you curious about what software they used to achieve these results. The secret to their success lies in the efficient use of different markup languages, with Markdown and \LaTeX \space being the most common choices for young researchers. In this section, we will approach the subject from a beginner’s perspective and assume that you have little to no prior knowledge of Markdown or \LaTeX \space. We will share some useful tips and tricks that we have used ourselves to learn markup languages, and we hope that you too can benefit from them.

\subsection{Markdown}
\label{Subsection: Markdown}

Markdown is a lightweight markup language that allows you to format text using an easy-to-read/write plain text format. A Markdown document can contain embedded graphics, source codes, and formulas. You may even have come across Markdown in several scenarios without realising it. Two of the most frequent encounters in our experience are Jupyter Notebook Markdown cells and GitHub README files. Markdown is also commonly used for academic blogs, online forums, collaborative software, and documentation pages. The entire history and usage of Markdown can be found on \href{https://en.wikipedia.org/wiki/Markdown}{Wikipedia}, so you don't really have to look too far! Plenty of Markdown cheat sheets can also be found via a quick online search.

Beginners often encounter challenges when attempting to include bullet points, equations, images or links within the default Markdown environment. However, alternatively, one could use third-party websites or software for mastering the basics of Markdown. One of the best `free' third-party websites for Markdown is called \href{https://hackmd.io}{HackMD}. HackMD offers many features, such as real-time collaboration, integration with other services such as Google Drive and GitHub, and notably, it also has a way of keeping track of any changes made. This is especially helpful where multiple people are working on a single document, thus allowing users to revert to previous versions if needed. However, its most valuable feature is the ability to render Markdown files whilst typing so that you can see exactly what's going on. Also, if perhaps you were previously formatting Markdown in a Jupyter Notebook, you can simply copy and paste everything from HackMD into the Jupyter Notebook Markdown cell. HackMD will allow you to sync contents to your GitHub repository if you are setting up your first repository. If you want to better visualize changes made to the README file, as you edit it, you can do so in HackMD and push these changes to your repository.

Alternatively, you can download one of the best Markdown software options (\emph{in my opinion}), such as VS code. With the help of some extensions, everything you were able to do using HackMD, can be accomplished using this platform. Although, for a beginner, we would advice sticking with HackMD as it is much easier to set up, and you can practise Markdown from pretty much anywhere. Thereafter, once you have gained more experience and feel more confident in switching between platforms, transitioning to VS Code will definitely benefit you in the long run.\\

There are numerous reasons to use Markdown, here are a few:
\begin{itemize}
    \item Relatively short learning curve.
    \item Being a lightweight markup language, it's texts are almost human-readable.
    \item Markdown is the best choice for a document to be published both on the Web and as printed text, as it can be easily converted to HTML.
\end{itemize}

\subsection{\LaTeX}
\LaTeX is a popular markup language for creating scientific documents/books/articles containing complex typographic elements such as Greek letters and mathematical expressions, and so you can appreciate their relevance in STEM.

To begin with, let’s discuss some of the mistakes we all made as beginners when we first picked up \LaTeX. I know my first mistake was wasting time trying to pick the best \LaTeX editor to use from the many options available, such as \href{www.overleaf.com}{Overleaf}, TeXmaker, TeXStudio and many others. All of these are great \LaTeX editors in their own right, but it is all too tempting to go down the rabbit hole of trying out all these editors and wanting to master all of them before ever learning how to use \LaTeX. Please learn from my mistakes, pick just one editor, and stick with it until you are comfortable using \LaTeX. Overleaf is the editor we most highly recommend for beginners (\textit{and what I am currently using to write this article}). A quick Google search will list all the advantages of using Overleaf, and that list will be fairly long. If you Google `How to use LaTeX’ (or a variation thereof), you will find tons of tutorials, documents and videos covering the subject. Overleaf has become one of the most widely accessed online-\LaTeX editors, offering a tutorial titled ` Learn LaTeX  in 30 minutes’. This detailed tutorial covers everything you need to know, from ‘What is LaTeX?’ to `Basic Formatting’ to `Downloading your finished documents’.

\textbf{Mistake \#2: creating your own template from scratch.} For experienced users, creating their own templates may still present some issues, thus its unlikely that a beginner will succeed right away, and doing so only adds to frustration, which eventually demotivates you from using \LaTeX. This is why we always recommend users use established templates as a starting point. Luckily, Overleaf is teeming with template options to help you get started, from curriculum vitaes (CVs), thesis templates, and even research paper templates that are specific to the journal you are submitting to. The Overleaf team have kindly organized everything into groups for you, depending on your needs, providing a user-friendly experience. Moreover, you have the convenience of searching for templates based on popular tags, recent uploads or even their selected, or, ‘featured’ templates. If there is ever something specific that you would like to change in your \LaTeX document, you can always refer to the Overleaf Help Library for more information. For example, if you would like to know how to insert hyperlinks in \LaTeX, just search ‘Hyperlink’. Suppose you are unsure about how to structure a complicated mathematical equation in \LaTeX or you are not sure if you structured your equations correctly, an online \LaTeX equation editor has proved to be rather handy.

Outside of writing reports, you may most likely come across \LaTeX used in academic presentations (\emph{probably why they look so much alike}). However, there's a reason why \LaTeX is so popular among academics for making presentations, despite their slightly `outdated' look. \LaTeX \space presentations, or, Beamer presentations offer a number of advantages over more traditional Microsoft PowerPoint, Keynote or Google Slides, and that is, the seamless incorporation of mathematical equations and code snippets, all using \LaTeX's  powerful typesetting capabilities. This makes all the difference for a computational scientist. Although, we note that one could also use equation tools (outside of Microsoft PowerPoint), and still get \LaTeX- quality formatting. Therefore, for those more comfortable with Keynote presentations (or variations thereof), one can still apply \LaTeX equation formatting and convert it to a high-quality png-rendered image using online \LaTeX equation editors such as `\LaTeX to png', or versions thereof.

\textbf{Looking for more reasons to use \LaTeX? Here are four of my favourite reasons:} 
\begin{itemize}
    \item Although the learning curve can be steep, once you become proficient, it is a more powerful tool than Markdown
    \item It is the right choice for complex and high-quality texts, especially for including mathematical or chemical equations.
    \item It is the gold standard in STEM regarding publishing scientific documents. Many journals accept raw \LaTeX documents for submission.
    \item There are tons of \LaTeX templates available online. Really! You could scroll through them for hours.
\end{itemize}

\section{Publishing (Python) Packages}\label{sec:tip3}

So you finally got your code working -- why not share it with the world? In this section I (Charlie Pilgrim) will discuss what I learnt from publishing a Python package \cite{pilgrim_piecewise-regression_2021} as a PhD student. Hopefully this will help you decide whether you want to publish your own package, and give you an idea of the benefits and challenges involved in doing so. While I will focus on Python packages, much of the advice here is general and will apply to any programming language. 

\subsection{Benefits}

The main benefits of publishing a package:

\begin{itemize}
    \item \textbf{Other people can benefit from your code!} This may not be a direct benefit to you, but it feels great to get an email from someone thanking you for your package which is saving them a lot of time and effort.
    \item \textbf{It looks great on your CV}. Having a published package is a strong signal that you are a competent programmer. This can be very helpful if you are going for any kind of position that requires coding experience, whether that be in industry or academia.  
    \item \textbf{Citations}. Often you can link a paper to the package, and ask people to cite the paper if they use the code. This could be a paper you have already written related to some research code (e.g. a new algorithm) or you can publish a short paper about the package itself. For example, the \textit{Journal of Open Source Software} is a venue for short articles that allow researchers to get cited for software.   
    \item \textbf{Get your name out there}. If you have developed some niche software then releasing the code is a good way to become known to researchers in your field. It is also perfectly reasonable to email other researchers in your area to ask them what they think of your software package - this can even lead to collaborations or job offers. 
    \item \textbf{Publishing code is a good practice to get into}. After publishing a package I found that I naturally wrote code that was tidier, better structured and generally closer to publishing. This kind of code is much easier to work with, especially when you come back to a project after a few months. And of course writing code to a high standard to begin with makes if easier to then publish the code in a public repository, which is becoming more important in research \cite{peng_reproducible_2011}. 
    \item \textbf{You will learn a lot}. As we will see in the next section, there are lots of challenges to publishing packages. While overcoming challenges can feel difficult at the time, it is a great way to learn.
\end{itemize}

\subsection{Challenges}

Mostly the challenges of publishing a package are around making your code \emph{production ready}. Software developers distinguish development code, which is essentially for your own use and as such can be messy and incomplete, and production code, which is for others to use and is of a generally higher quality. 

Getting your code production ready and published will involve:

\begin{itemize}
    \item \textbf{Tidying up}. As you develop code there is a tendency for things to get messy and confusing. Before publishing any code you will probably want to spend some time and effort tidying things up.  
    \item \textbf{Writing Documentation}. You will need to tell users how to use your code. This includes the README file and other written documentation, as well as elements within the code itself such as function definitions and code comments.
    \item \textbf{Testing}. While writing your software you probably did at least some manual testing to check a few things work as expected. In order to be fully confident that the package works as expected you might want to write some more standardised tests such as \emph{unit} or \emph{functional} tests. 
    \item \textbf{Covering edge cases}. What happens if a user does something unexpected? This may or may not be important to cover, depending on who you expect to use the package. 
    \item \textbf{Extra features}. You will feel pressure to add extra functionality. Try and resist this! Doing one thing very well is a great aim for a package. 
    \item \textbf{Actually publishing the package}. This is actually one of the simpler steps! For example, there are a number of tutorials to publish code to the Python package index which take less than an hour. 
    \item \textbf{Ongoing maintenance}. You may get requests for new features, or people might tell you about bugs. And you might need to update your package to work with new versions of software dependencies as they are released e.g. new Python versions. Of course, you can decide whether to address these issues when they arise.  
    \item \textbf{Intellectual property and licensing}. When you publish code you will also need to provide a license that sets out the rules for how other people can use the code. A common choice is permissive open-source licenses such as the MIT license which allow anyone to re-use the code. If you are using or adapting someone else's code \cite{Pilgrim2023Apr} then it is good practice to make sure that you follow the rules set out in any existing licenses.  
\end{itemize}

\subsection{Conclusion}

Only you can decide whether to publish a package. Is it worth it? If you want to publish a package for general consumption then there is a lot of work involved in getting the code \emph{production ready}. This work may be worth it if it will really help your career or you have a burning desire to publish the code. However in academia you might want to publish a small niche package that does just one thing very well. This could be extremely helpful for a small number of researchers. Often this kind of package does not need to be quite as polished as a more general package, as the users will be willing to work through any issues. 

Hopefully this article has made the whole process of publishing packages a little more transparent. Even if you don't personally publish a package, it is generally useful to have an idea of what is involved in the process.

\section{How Do I Make My Work Reproducible?}

Science, in general, has been plagued by results that are not easily reproducible - and perhaps to some extent, this is to be expected: cutting-edge research is inherently difficult; if it weren't, it probably wouldn't be cutting edge. However, we should also expect it to be easier on the second or third attempt.

Reproducibility is an essential component of scientific research. It is what ensures the credibility of results discussed in research journals. What good is it if a research group at one institute claims the synthesis of a previously undiscovered material, but another research group, at another university cannot follow their experimental methodology to produce the same ``novel'' material. Well, the same is true of computational code. In the ensuing discussion, we highlight seven key points that we deem most important to keep in mind when publishing code, so that one might be able to replicate the results provided.

\begin{enumerate}
    \item \textbf{Documentation:} Perhaps an obvious point, but detailed documentation goes a long way. This must include a clear description of how your code is organised, what each script does, and how to execute the code. Popular tools for this include the likes of ReadtheDocs\cite{readthedocs} and Doxygen.\cite{doxygen}
    
    \item \textbf{Version control:} In the previous section we discussed how one might publish a Python package. This will have likely involved keeping track of previous versions of code using the likes of Git.\cite{git} Platforms like Gitlab\cite{gitlab} and GitHub\cite{github} are primed for this sort of thing, and are indispensable tools for keeping up to date with the latest versions of code.
    
    \item \textbf{Code dependencies:} Declare all code dependencies and versions of software and libraries used from the outset. You might typically include this in the README.md file as part of your code package. Also, the use of virtual environments can help you to avoid any clashes. If you have used random number seeds anywhere in your code, you should clearly document that, as this is critical for reproducibility.
    
    \item \textbf{Tutorials:} Again, as with documentation, having a full tutorial of the code e.g., in the form of a Jupyter Notebook, complete with visualisation of expected results, will be extremely helpful to users, so that they know what to expect.
    
    \item \textbf{Open source:} Perhaps another obvious point, but any data used to produce your results should be made publicly available to the users of your code. This can be easily done through platform like Figshare and Zenodo. All it takes is citing this link in your research paper.  
    
    \item \textbf{Unit testing:} As discussed in the previous section, it might help if you include some sort of unit testing infrastructure in your published code e.g., \texttt{Pytest}. This will help to ensure the accuracy of your code.
    
    \item \textbf{General good coding practices:} By sticking to traditional coding standards you reduce the risk of users misinterpreting your code, and thereby are more likely to have success in replicating your results. This includes things like how you structure your code, e.g., keeping the \texttt{src}, \texttt{examples}, and \texttt{docs} directories separate. Obvious! but necessary if you want to avoid confusion.
\end{enumerate}

\section{Recommended Software and Resources}

For an impressive list of software programs for computational chemistry and physics, check out \href{https://silicostudio.com/unlock-the-power-of-computational-chemistry/}{silicostudio} by David Abassi.\cite{SilicoStudio} Additionally, below are a few other websites, papers, and YouTube channels that you may find helpful.\\

\noindent \textbf{Websites}

\begin{itemize}
    \item MolSSI best practices for software development.\cite{Molssi} - a teaching resource by The  molecular sciences software institute
    \item The missing semester of your computer science degree\cite{Missingsemester} - a seminar series by graduate students at MIT.
    \item Introduction to Deep Learning.\cite{DeepLearningMIT} - a lecture course by graduate students at MIT.
\end{itemize}

\noindent \textbf{YouTube Channels}

\begin{itemize}
    \item TMP Chem\cite{tmp} - introductory videos on computational chemistry. 
    \item Virtual Simulation Lab\cite{vsl}- computational tools seminar series introducing version control, visualization, uncertainty quantification, and much more.
    \item StatQuest\cite{statquest}- videos on statistics, machine learning, and data science. 
    \item The Computational Toolkit\cite{comptoolkit}- a virtual seminar series offering a practical introduction to tools, tips, and tricks in computational science.
\end{itemize}

\textbf{Papers and other resources
}
\begin{itemize}
    \item \citet{CumbyJOSE2023} - a collection of Jupyter Notebooks to guide students with data-driven chemistry, and introduces Python programming along with its usage in data analysis, typically required for a Chemistry degree.
    \item \citet{FrenchJOSE2023} - a sequential learning system designed to guide the reader on how to use R programming to analyze data.
    \item \citet{JuliaDataScience} - an open source and open access book on how to do data science using the Julia programming language.
    \item \citet{Pilgrim2023Apr} - ten simple rules for working with other people's code. 
    \item \citet{white2021deep} - an ebook on deep learning for molecules and materials.
\end{itemize}

\section{Conclusions}
In this paper, we have identified and discussed various topics that we believe will be useful and relevant to academic researchers starting a career in computational science. Here, we have advised where new researchers can look to get started with their research, covered useful programming languages that will likely be required at some point, as well as other key topics that are frequently used such as machine learning and high-performance computing. We also discuss some pastoral aspects associated with computational research and encourage students to discuss matters with other academics, both within and outside their research group. This paper should act as a useful resource to new computational scientists and aid them as they start their research careers, as well as learn about important concepts encountered by previous computational science PhD students.

\section*{Acknowledgements}
I.I., D.M., J.M.T., Z.F., C.M., and C.D.W. acknowledge funding from the EPSRC Centre for Doctoral Training in Modelling of Heterogeneous Systems [EP/S022848/1]. S.C. acknowledges funding from the EPSRC Centre for Doctoral Training in Diamond Science and Technology [EP/L015315/1] and the Research Development Fund of the University of Warwick. C.P. acknowledges funding from the EPSRC Mathematics for Real-World Systems Centre for Doctoral Training [EP/S022244/1]. In addition, we would like to acknowledge the valuable contributions of several other colleagues for their early efforts and input in the computational toolkit seminar series, on which this paper is based, and they are; Peter Lewin-Jones, Kyle Fogarty, Lakshmi Shenoy, Matthew Harrison \& Charlotte Rogerson. Finally, we would like to thank both Professors James Kermode and Julie Staunton for their time in reading our drafts and offering valuable advice and comments on our manuscript.



\providecommand{\latin}[1]{#1}
\makeatletter
\providecommand{\doi}
  {\begingroup\let\do\@makeother\dospecials
  \catcode`\{=1 \catcode`\}=2 \doi@aux}
\providecommand{\doi@aux}[1]{\endgroup\texttt{#1}}
\makeatother
\providecommand*\mcitethebibliography{\thebibliography}
\csname @ifundefined\endcsname{endmcitethebibliography}
  {\let\endmcitethebibliography\endthebibliography}{}
\begin{mcitethebibliography}{51}
\providecommand*\natexlab[1]{#1}
\providecommand*\mciteSetBstSublistMode[1]{}
\providecommand*\mciteSetBstMaxWidthForm[2]{}
\providecommand*\mciteBstWouldAddEndPuncttrue
  {\def\EndOfBibitem{\unskip.}}
\providecommand*\mciteBstWouldAddEndPunctfalse
  {\let\EndOfBibitem\relax}
\providecommand*\mciteSetBstMidEndSepPunct[3]{}
\providecommand*\mciteSetBstSublistLabelBeginEnd[3]{}
\providecommand*\EndOfBibitem{}
\mciteSetBstSublistMode{f}
\mciteSetBstMaxWidthForm{subitem}{(\alph{mcitesubitemcount})}
\mciteSetBstSublistLabelBeginEnd
  {\mcitemaxwidthsubitemform\space}
  {\relax}
  {\relax}

\bibitem[Clark \latin{et~al.}(2005)Clark, Segall, Pickard, Hasnip, Probert,
  Refson, and Payne]{CASTEP}
Clark,~S.~J.; Segall,~M.~D.; Pickard,~C.~J.; Hasnip,~P.~J.; Probert,~M.~J.;
  Refson,~K.; Payne,~M. First principles methods using {CASTEP}. \emph{Z.
  Kristall.} \textbf{2005}, \emph{220}, 567--570\relax
\mciteBstWouldAddEndPuncttrue
\mciteSetBstMidEndSepPunct{\mcitedefaultmidpunct}
{\mcitedefaultendpunct}{\mcitedefaultseppunct}\relax
\EndOfBibitem
\bibitem[Rüßmann \latin{et~al.}(2022)Rüßmann, Mavropoulos, Zeller, Bouaziz,
  Dos Santos~Dias, Blügel, Bauer, Baumeister, Bornemann, Brinker, Dederichs,
  Drittler, Essing, Géranton, Long, Lounis, Mendive~Tapia, Rabel, Dos~Santos,
  Schweflinghaus, Antognini~Silva, Thiess, and Zimmermann]{JuKKR}
Rüßmann,~P. \latin{et~al.}  The JuKKR code. 2022\relax
\mciteBstWouldAddEndPuncttrue
\mciteSetBstMidEndSepPunct{\mcitedefaultmidpunct}
{\mcitedefaultendpunct}{\mcitedefaultseppunct}\relax
\EndOfBibitem
\bibitem[Woodgate \latin{et~al.}(2023)Woodgate, Hedlund, Lewis, and
  Staunton]{woodgate_interplay_2023}
Woodgate,~C.~D.; Hedlund,~D.; Lewis,~L.~H.; Staunton,~J.~B. Interplay between
  magnetism and short-range order in medium- and high-entropy alloys: CrCoNi,
  CrFeCoNi, and CrMnFeCoNi. \emph{Phys. Rev. Mater.} \textbf{2023}, \emph{7},
  053801\relax
\mciteBstWouldAddEndPuncttrue
\mciteSetBstMidEndSepPunct{\mcitedefaultmidpunct}
{\mcitedefaultendpunct}{\mcitedefaultseppunct}\relax
\EndOfBibitem
\bibitem[Chadwick(2023)]{linux_ryan}
Chadwick,~R. {Linux Tutorial for Beginners - Learn Linux and the Bash Command
  Line}. 2023; \url{https://ryanstutorials.net/linuxtutorial}, [Online;
  accessed 20. Oct. 2023]\relax
\mciteBstWouldAddEndPuncttrue
\mciteSetBstMidEndSepPunct{\mcitedefaultmidpunct}
{\mcitedefaultendpunct}{\mcitedefaultseppunct}\relax
\EndOfBibitem
\bibitem[Chadwick(2023)]{bash_ryan}
Chadwick,~R. {Bash Scripting Tutorial - Ryans Tutorials}. 2023;
  \url{https://ryanstutorials.net/bash-scripting-tutorial}, [Online; accessed
  20. Oct. 2023]\relax
\mciteBstWouldAddEndPuncttrue
\mciteSetBstMidEndSepPunct{\mcitedefaultmidpunct}
{\mcitedefaultendpunct}{\mcitedefaultseppunct}\relax
\EndOfBibitem
\bibitem[Dowd and Severance(1998)Dowd, and Severance]{Dowd98}
Dowd,~K.; Severance,~C.~R. \emph{\textnormal{High performance computing}};
  O'Reilly \& Associates, Cambridge, 1998\relax
\mciteBstWouldAddEndPuncttrue
\mciteSetBstMidEndSepPunct{\mcitedefaultmidpunct}
{\mcitedefaultendpunct}{\mcitedefaultseppunct}\relax
\EndOfBibitem
\bibitem[Huhn \latin{et~al.}(2020)Huhn, Lange, Yu, Yoon, and Blum]{HuhnCPC20}
Huhn,~W.~P.; Lange,~B.; Yu,~V. W.-z.; Yoon,~M.; Blum,~V. {GPU} acceleration of
  all-electron electronic structure theory using localized numeric
  atom-centered basis functions. \emph{Comput. Phys. Commun.} \textbf{2020},
  \emph{254}, 107314\relax
\mciteBstWouldAddEndPuncttrue
\mciteSetBstMidEndSepPunct{\mcitedefaultmidpunct}
{\mcitedefaultendpunct}{\mcitedefaultseppunct}\relax
\EndOfBibitem
\bibitem[Spiga and Girotto(2012)Spiga, and Girotto]{SpigaIEEE12}
Spiga,~F.; Girotto,~I. {phiGEMM}: {A} {CPU}-{GPU} {Library} for {Porting}
  {Quantum} {ESPRESSO} on {Hybrid} {Systems}. 2012 20th {Euromicro}
  {International} {Conference} on {Parallel}, {Distributed} and {Network}-based
  {Processing}. 2012; pp 368--375, ISSN: 2377-5750\relax
\mciteBstWouldAddEndPuncttrue
\mciteSetBstMidEndSepPunct{\mcitedefaultmidpunct}
{\mcitedefaultendpunct}{\mcitedefaultseppunct}\relax
\EndOfBibitem
\bibitem[Vogt \latin{et~al.}(2008)Vogt, Olivares-Amaya, Kermes, Shao,
  Amador-Bedolla, and Aspuru-Guzik]{VogtJPCA08}
Vogt,~L.; Olivares-Amaya,~R.; Kermes,~S.; Shao,~Y.; Amador-Bedolla,~C.;
  Aspuru-Guzik,~A. Accelerating {Resolution}-of-the-{Identity} {Second}-{Order}
  {Møller}-{Plesset} {Quantum} {Chemistry} {Calculations} with {Graphical}
  {Processing} {Units}. \emph{J. Phys. Chem. A} \textbf{2008}, \emph{112},
  2049--2057\relax
\mciteBstWouldAddEndPuncttrue
\mciteSetBstMidEndSepPunct{\mcitedefaultmidpunct}
{\mcitedefaultendpunct}{\mcitedefaultseppunct}\relax
\EndOfBibitem
\bibitem[Wilkinson and Skylaris(2013)Wilkinson, and Skylaris]{WilkinsonJCC13}
Wilkinson,~K.; Skylaris,~C.-K. Porting {ONETEP} to graphical processing
  unit-based coprocessors. 1. {FFT} box operations. \emph{J. Comp. Chem.}
  \textbf{2013}, \emph{34}, 2446--2459\relax
\mciteBstWouldAddEndPuncttrue
\mciteSetBstMidEndSepPunct{\mcitedefaultmidpunct}
{\mcitedefaultendpunct}{\mcitedefaultseppunct}\relax
\EndOfBibitem
\bibitem[Yan \latin{et~al.}(2013)Yan, Li, and O’Grady]{YanCPC13}
Yan,~J.; Li,~L.; O’Grady,~C. Graphics {Processing} {Unit} acceleration of the
  {Random} {Phase} {Approximation} in the projector augmented wave method.
  \emph{Comput. Phys. Commun.} \textbf{2013}, \emph{184}, 2728--2733\relax
\mciteBstWouldAddEndPuncttrue
\mciteSetBstMidEndSepPunct{\mcitedefaultmidpunct}
{\mcitedefaultendpunct}{\mcitedefaultseppunct}\relax
\EndOfBibitem
\bibitem[Genovese \latin{et~al.}(2009)Genovese, Ospici, Deutsch, Méhaut,
  Neelov, and Goedecker]{GenoveseJCP09}
Genovese,~L.; Ospici,~M.; Deutsch,~T.; Méhaut,~J.-F.; Neelov,~A.;
  Goedecker,~S. Density functional theory calculation on many-cores hybrid
  central processing unit-graphic processing unit architectures. \emph{J. Chem.
  Phys.} \textbf{2009}, \emph{131}, 034103\relax
\mciteBstWouldAddEndPuncttrue
\mciteSetBstMidEndSepPunct{\mcitedefaultmidpunct}
{\mcitedefaultendpunct}{\mcitedefaultseppunct}\relax
\EndOfBibitem
\bibitem[Bishop(2006)]{Bishop2006Jan}
Bishop,~C. {Pattern Recognition and Machine Learning}. \emph{J. Electron.
  Imaging} \textbf{2006}, \emph{16}, 140--155\relax
\mciteBstWouldAddEndPuncttrue
\mciteSetBstMidEndSepPunct{\mcitedefaultmidpunct}
{\mcitedefaultendpunct}{\mcitedefaultseppunct}\relax
\EndOfBibitem
\bibitem[Murphy and Bach(2012)Murphy, and Bach]{Murphy2012Sep}
Murphy,~K.~P.; Bach,~F. \emph{{Machine Learning: A Probabilistic Perspective
  (Adaptive Computation and Machine Learning Series)}}; MIT Press: London,
  England, UK, 2012\relax
\mciteBstWouldAddEndPuncttrue
\mciteSetBstMidEndSepPunct{\mcitedefaultmidpunct}
{\mcitedefaultendpunct}{\mcitedefaultseppunct}\relax
\EndOfBibitem
\bibitem[Lones(2021)]{lones_how_2021}
Lones,~M.~A. How to avoid machine learning pitfalls: a guide for academic
  researchers. \emph{arXiv:2108.02497 [cs]} \textbf{2021}, arXiv:
  2108.02497\relax
\mciteBstWouldAddEndPuncttrue
\mciteSetBstMidEndSepPunct{\mcitedefaultmidpunct}
{\mcitedefaultendpunct}{\mcitedefaultseppunct}\relax
\EndOfBibitem
\bibitem[Belyaev \latin{et~al.}(2014)Belyaev, Burnaev, and
  Kapushev]{Belyaev2014Mar}
Belyaev,~M.; Burnaev,~E.; Kapushev,~Y. {Exact Inference for Gaussian Process
  Regression in case of Big Data with the Cartesian Product Structure}.
  \emph{arXiv} \textbf{2014}, \relax
\mciteBstWouldAddEndPunctfalse
\mciteSetBstMidEndSepPunct{\mcitedefaultmidpunct}
{}{\mcitedefaultseppunct}\relax
\EndOfBibitem
\bibitem[Liu \latin{et~al.}(2018)Liu, Ong, Shen, and Cai]{Liu2018Jul}
Liu,~H.; Ong,~Y.-S.; Shen,~X.; Cai,~J. {When Gaussian Process Meets Big Data: A
  Review of Scalable GPs}. \emph{arXiv} \textbf{2018}, \relax
\mciteBstWouldAddEndPunctfalse
\mciteSetBstMidEndSepPunct{\mcitedefaultmidpunct}
{}{\mcitedefaultseppunct}\relax
\EndOfBibitem
\bibitem[Buitinck \latin{et~al.}(2013)Buitinck, Louppe, Blondel, Pedregosa,
  Mueller, Grisel, Niculae, Prettenhofer, Gramfort, Grobler, Layton,
  VanderPlas, Joly, Holt, and Varoquaux]{sklearn_api}
Buitinck,~L.; Louppe,~G.; Blondel,~M.; Pedregosa,~F.; Mueller,~A.; Grisel,~O.;
  Niculae,~V.; Prettenhofer,~P.; Gramfort,~A.; Grobler,~J.; Layton,~R.;
  VanderPlas,~J.; Joly,~A.; Holt,~B.; Varoquaux,~G. {API} design for machine
  learning software: experiences from the scikit-learn project. ECML PKDD
  Workshop: Languages for Data Mining and Machine Learning. 2013; pp
  108--122\relax
\mciteBstWouldAddEndPuncttrue
\mciteSetBstMidEndSepPunct{\mcitedefaultmidpunct}
{\mcitedefaultendpunct}{\mcitedefaultseppunct}\relax
\EndOfBibitem
\bibitem[Pedregosa \latin{et~al.}(2011)Pedregosa, Varoquaux, Gramfort, Michel,
  Thirion, Grisel, Blondel, Prettenhofer, Weiss, Dubourg, Vanderplas, Passos,
  Cournapeau, Brucher, Perrot, and Duchesnay]{Scikit-learn}
Pedregosa,~F. \latin{et~al.}  Scikit-learn: Machine Learning in {P}ython.
  \emph{J. Mach. Learn. Res.} \textbf{2011}, \emph{12}, 2825--2830\relax
\mciteBstWouldAddEndPuncttrue
\mciteSetBstMidEndSepPunct{\mcitedefaultmidpunct}
{\mcitedefaultendpunct}{\mcitedefaultseppunct}\relax
\EndOfBibitem
\bibitem[Abadi \latin{et~al.}(2015)Abadi, Agarwal, Barham, Brevdo, Chen, Citro,
  Corrado, Davis, Dean, Devin, Ghemawat, Goodfellow, Harp, Irving, Isard, Jia,
  Jozefowicz, Kaiser, Kudlur, Levenberg, Man\'{e}, Monga, Moore, Murray, Olah,
  Schuster, Shlens, Steiner, Sutskever, Talwar, Tucker, Vanhoucke, Vasudevan,
  Vi\'{e}gas, Vinyals, Warden, Wattenberg, Wicke, Yu, and Zheng]{TensorFlow}
Abadi,~M. \latin{et~al.}  TensorFlow: Large-Scale Machine Learning on
  Heterogeneous Systems. 2015; \url{https://www.tensorflow.org/}, Software
  available from tensorflow.org\relax
\mciteBstWouldAddEndPuncttrue
\mciteSetBstMidEndSepPunct{\mcitedefaultmidpunct}
{\mcitedefaultendpunct}{\mcitedefaultseppunct}\relax
\EndOfBibitem
\bibitem[Paszke \latin{et~al.}(2019)Paszke, Gross, Massa, Lerer, Bradbury,
  Chanan, Killeen, Lin, Gimelshein, Antiga, Desmaison, Kopf, Yang, DeVito,
  Raison, Tejani, Chilamkurthy, Steiner, Fang, Bai, and Chintala]{PyTorch}
Paszke,~A. \latin{et~al.}  \emph{Advances in Neural Information Processing
  Systems 32}; Curran Associates, Inc., 2019; pp 8024--8035\relax
\mciteBstWouldAddEndPuncttrue
\mciteSetBstMidEndSepPunct{\mcitedefaultmidpunct}
{\mcitedefaultendpunct}{\mcitedefaultseppunct}\relax
\EndOfBibitem
\bibitem[Hunter(2007)]{Matplotlib}
Hunter,~J.~D. {Matplotlib: A 2D Graphics Environment}. \emph{Comput. Sci. Eng.}
  \textbf{2007}, \emph{3}, 90--95\relax
\mciteBstWouldAddEndPuncttrue
\mciteSetBstMidEndSepPunct{\mcitedefaultmidpunct}
{\mcitedefaultendpunct}{\mcitedefaultseppunct}\relax
\EndOfBibitem
\bibitem[Harris \latin{et~al.}(2007)Harris, Jarrod~Millman, van~der Walt,
  Gommers, Virtanen, Cournapeau, Wieser, Taylor, Berg, Smith, Kern, Picus,
  Hoyer, van Kerkwijk, Brett, Haldane, Fern{\'a}ndez~del R{\'i}o, Wiebe,
  Peterson, G{\'e}rard-Marchant, Sheppard, Reddy, Weckesser, Abbasi, Gohlke,
  and Oliphant]{NumPy}
Harris,~C.~R. \latin{et~al.}  {Array programming with NumPy}. \emph{Nature}
  \textbf{2007}, \emph{585}, 357--362\relax
\mciteBstWouldAddEndPuncttrue
\mciteSetBstMidEndSepPunct{\mcitedefaultmidpunct}
{\mcitedefaultendpunct}{\mcitedefaultseppunct}\relax
\EndOfBibitem
\bibitem[Virtanen \latin{et~al.}(2020)Virtanen, Gommers, Oliphant, Haberland,
  Reddy, Cournapeau, Burovski, Peterson, Weckesser, Bright, van~der Walt,
  Brett, Wilson, Jarod~Millman, Mayorov, Nelson, Jones, Kern, Larson, Carey,
  Polat, Feng, Moore, VanderPlas, Laxalde, Perktold, Cimrman, Heriksen,
  Quintero, Harris, Archibald, Ribeiro, Pedregosa, van Mulbregt, and {SciPy 1.0
  Contributors}]{SciPy}
Virtanen,~P. \latin{et~al.}  {SciPy 1.0: fundamental algorithms for scientific
  computing in Python}. \emph{Nature Methods} \textbf{2020}, \emph{17},
  261--272\relax
\mciteBstWouldAddEndPuncttrue
\mciteSetBstMidEndSepPunct{\mcitedefaultmidpunct}
{\mcitedefaultendpunct}{\mcitedefaultseppunct}\relax
\EndOfBibitem
\bibitem[Waskom(2021)]{Seaborn}
Waskom,~M.~L. seaborn: statistical data visualization. \emph{J. Open Source
  Softw.} \textbf{2021}, \emph{6}, 3021\relax
\mciteBstWouldAddEndPuncttrue
\mciteSetBstMidEndSepPunct{\mcitedefaultmidpunct}
{\mcitedefaultendpunct}{\mcitedefaultseppunct}\relax
\EndOfBibitem
\bibitem[pandas~development team(2020)]{Pandas_Zenodo}
pandas~development team,~T. pandas-dev/pandas: Pandas. 2020;
  \url{https://doi.org/10.5281/zenodo.3509134}\relax
\mciteBstWouldAddEndPuncttrue
\mciteSetBstMidEndSepPunct{\mcitedefaultmidpunct}
{\mcitedefaultendpunct}{\mcitedefaultseppunct}\relax
\EndOfBibitem
\bibitem[Kluyver \latin{et~al.}(2016)Kluyver, Ragan-Kelley, P{\'e}rez, Granger,
  Bussonnier, Frederic, Kelley, Hamrick, Grout, Corlay, Ivanov, Avila, Abdalla,
  and Willing]{Jupyter}
Kluyver,~T.; Ragan-Kelley,~B.; P{\'e}rez,~F.; Granger,~B.; Bussonnier,~M.;
  Frederic,~J.; Kelley,~K.; Hamrick,~J.; Grout,~J.; Corlay,~S.; Ivanov,~P.;
  Avila,~D.; Abdalla,~S.; Willing,~C. Jupyter Notebooks -- a publishing format
  for reproducible computational workflows. Positioning and Power in Academic
  Publishing: Players, Agents and Agendas. 2016; pp 87 -- 90\relax
\mciteBstWouldAddEndPuncttrue
\mciteSetBstMidEndSepPunct{\mcitedefaultmidpunct}
{\mcitedefaultendpunct}{\mcitedefaultseppunct}\relax
\EndOfBibitem
\bibitem[Abbott \latin{et~al.}(2020)Abbott, Abbott, Abbott, Abraham, Acernese,
  Ackley, Adams, Adya, Affeldt, Agathos, Agatsuma, Aggarwal, Aguiar, Aiello,
  Ain, Ajith, Alford, Allen, Allocca, Aloy, Altin, Amato, Ananyeva, Anderson,
  Anderson, Angelova, Antier, Appert, Arai, Araya, Areeda,
  Ar{\ifmmode\grave{e}\else\`{e}\fi}ne, Arnaud, Arun, Ascenzi, Ashton, Aston,
  Astone, Aubin, Aufmuth, AultONeal, Austin, Avendano, Avila-Alvarez, Babak,
  Bacon, Badaracco, Bader, Bae, Baker, Baldaccini, Ballardin, Ballmer,
  Banagiri, Barayoga, Barclay, Barish, Barker, Barkett, Barnum, Barone, Barr,
  Barsotti, Barsuglia, Barta, Bartlett, Bartos, Bassiri, Basti, Bawaj, Bayley,
  Bazzan, B{\ifmmode\acute{e}\else\'{e}\fi}csy, Bejger, Belahcene, Bell,
  Beniwal, Berger, Bergmann, Bernuzzi, Bero, Berry, Bersanetti, Bertolini,
  Betzwieser, Bhandare, Bidler, Bilenko, Bilgili, Billingsley, Birch, Birney,
  Birnholtz, Biscans, Biscoveanu, Bisht, Bitossi, Bizouard, Blackburn, Blair,
  Blair, Blair, Bloemen, Bode, Boer, Boetzel, Bogaert, Bondu, Bonilla, Bonnand,
  Booker, Boom, Booth, Bork, Boschi, Bose, Bossie, Bossilkov, Bosveld,
  Bouffanais, Bozzi, Bradaschia, Brady, Bramley, Branchesi, Brau, Briant,
  Briggs, Brighenti, Brillet, Brinkmann, Brisson, Brockill, Brooks, Brown,
  Brunett, Buikema, Bulik, Bulten, Buonanno, Buskulic, Buy, Byer, Cabero,
  Cadonati, Cagnoli, Cahillane, Bustillo, Callister, Calloni, Camp, Campbell,
  Canepa, Cannon, Cao, Cao, Capocasa, Carbognani, Caride, Carney, Carullo,
  Diaz, Casentini, Caudill, Cavagli{\ifmmode\grave{a}\else\`{a}\fi}, Cavalier,
  Cavalieri, Cella,
  Cerd{\ifmmode\acute{a}\else\'{a}\fi}-Dur{\ifmmode\acute{a}\else\'{a}\fi}n,
  Cerretani, Cesarini, Chaibi, Chakravarti, Chamberlin, Chan, Chao, Charlton,
  Chase, Chassande-Mottin, Chatterjee, Chaturvedi, Chatziioannou, Cheeseboro,
  Chen, Chen, Chen, Cheng, Cheong, Chia, Chincarini, Chiummo, Cho, Cho, Cho,
  Christensen, Chu, Chua, Chung, Chung, Ciani, Ciobanu, Ciolfi, Cipriano,
  Cirone, Clara, Clark, Clearwater, Cleva, Cocchieri, Coccia, Cohadon, Cohen,
  Colgan, Colleoni, Collette, Collins, Cominsky, Constancio, Conti, Cooper,
  Corban, Corbitt, Cordero-Carri{\ifmmode\acute{o}\else\'{o}\fi}n, Corley,
  Cornish, Corsi, Cortese, Costa, Cotesta, Coughlin, Coughlin, Coulon,
  Countryman, Couvares, Covas, Cowan, Coward, Cowart, Coyne, Coyne, Creighton,
  Creighton, Cripe, Croquette, Crowder, Cullen, Cumming, Cunningham, Cuoco,
  Dal~Canton, D{\ifmmode\acute{a}\else\'{a}\fi}lya, Danilishin, D{'}Antonio,
  Danzmann, Dasgupta, Da~Silva~Costa, Datrier, Dattilo, Dave, Davier, Davis,
  Daw, DeBra, Deenadayalan, Degallaix, De~Laurentis,
  Del{\ifmmode\acute{e}\else\'{e}\fi}glise, Del~Pozzo, DeMarchi, Demos, Dent,
  De~Pietri, Derby, De~Rosa, De~Rossi, DeSalvo, de~Varona, Dhurandhar,
  D{\ifmmode\acute{\imath}\else\'{\i}\fi}az, Dietrich, Di~Fiore, Di~Giovanni,
  Di~Girolamo, Di~Lieto, Ding, Di~Pace, Di~Palma, Di~Renzo, Dmitriev, Doctor,
  Donovan, Dooley, Doravari, Dorrington, Downes, Drago, Driggers, Du, Ducoin,
  Dupej, Dwyer, Easter, Edo, Edwards, Effler, Ehrens, Eichholz, Eikenberry,
  Eisenmann, Eisenstein, Essick, Estelles, Estevez, Etienne, Etzel, Evans,
  Evans, Fafone, Fair, Fairhurst, Fan, Farinon, Farr, Farr, Fauchon-Jones,
  Favata, Fays, Fazio, Fee, Feicht, Fejer, Feng, Fernandez-Galiana, Ferrante,
  Ferreira, Ferreira, Ferrini, Fidecaro, Fiori, Fiorucci, Fishbach, Fisher,
  Fishner, Fitz-Axen, Flaminio, Fletcher, Flynn, Fong, Font, Forsyth, Fournier,
  Frasca, Frasconi, Frei, Freise, Frey, Frey, Fritschel, Frolov, Fulda, Fyffe,
  Gabbard, Gadre, Gaebel, Gair, Gammaitoni, Ganija, Gaonkar, Garcia,
  Garc{\ifmmode\acute{\imath}\else\'{\i}\fi}a-Quir{\ifmmode\acute{o}\else\'{o}\fi}s,
  Garufi, Gateley, Gaudio, Gaur, Gayathri, Gemme, Genin, Gennai, George,
  George, Gergely, Germain, Ghonge, Ghosh, Ghosh, Ghosh, Giacomazzo, Giaime,
  Giardina, Giazotto, Gill, Giordano, Glover, Godwin, Goetz, Goetz, Goncharov,
  Gonz{\ifmmode\acute{a}\else\'{a}\fi}lez, Castro, Gopakumar, Gorodetsky,
  Gossan, Gosselin, Gouaty, Grado, Graef, Granata, Grant, Gras, Grassia, Gray,
  Gray, Greco, Green, Green, Gretarsson, Groot, Grote, Grunewald, Gruning,
  Guidi, Gulati, Guo, Gupta, Gupta, Gustafson, Gustafson, Haegel, Halim, Hall,
  Hall, Hamilton, Hammond, Haney, Hanke, Hanks, Hanna, Hannam, Hannuksela,
  Hanson, Hardwick, Haris, Harms, Harry, Harry, Haster, Haughian, Hayes, Healy,
  Heidmann, Heintze, Heitmann, Hello, Hemming, Hendry, Heng, Hennig,
  Heptonstall, Vivanco, Heurs, Hild, Hinderer, Hoak, Hochheim, Hofman, Holgado,
  Holland, Holt, Holz, Hopkins, Horst, Hough, Howell, Hoy, Hreibi, Huerta,
  Huet, Hughey, Hulko, Husa, Huttner, Huynh-Dinh, Idzkowski, Iess, Ingram,
  Inta, Intini, Irwin, Isa, Isac, Isi, Iyer, Izumi, Jacqmin, Jadhav, Jani,
  Janthalur, Jaranowski, Jenkins, Jiang, Johnson, Jones, Jones, Jones, Jonker,
  Ju, Junker, Kalaghatgi, Kalogera, Kamai, Kandhasamy, Kang, Kanner, Kapadia,
  Karki, Karvinen, Kashyap, Kasprzack, Katsanevas, Katsavounidis, Katzman,
  Kaufer, Kawabe, Keerthana,
  K{\ifmmode\acute{e}\else\'{e}\fi}f{\ifmmode\acute{e}\else\'{e}\fi}lian,
  Keitel, Kennedy, Key, Khalili, Khan, Khan, Khan, Khan, Khazanov, Khursheed,
  Kijbunchoo, Kim, Kim, Kim, Kim, Kim, Kim, Kim, Kimball, King, King,
  Kinley-Hanlon, Kirchhoff, Kissel, Kleybolte, Klika, Klimenko, Knowles, Koch,
  Koehlenbeck, Koekoek, Koley, Kondrashov, Kontos, Koper, Korobko, Korth,
  Kowalska, Kozak, Kringel, Krishnendu, Kr{\ifmmode\acute{o}\else\'{o}\fi}lak,
  Kuehn, Kumar, Kumar, Kumar, Kumar, Kuo, Kutynia, Kwang, Lackey, Lai, Lam,
  Landry, Lane, Lang, Lange, Lantz, Lanza, Larson, Lartaux-Vollard, Lasky,
  Laxen, Lazzarini, Lazzaro, Leaci, Leavey, Lecoeuche, Lee, Lee, Lee, Lee, Lee,
  Lee, Lehmann, Lenon, Leroy, Letendre, Levin, Li, Li, Li, Li, Lin, Linde,
  Linker, Littenberg, Liu, Liu, Lo, Lockerbie, London, Longo, Lorenzini,
  Loriette, Lormand, Losurdo, Lough, Lousto, Lovelace, Lower,
  L{\ifmmode\ddot{u}\else\"{u}\fi}ck, Lumaca, Lundgren, Lynch, Ma, Macas,
  Macfoy, MacInnis, Macleod, Macquet,
  Maga{\ifmmode\tilde{n}\else\~{n}\fi}a-Sandoval, Zertuche, Magee, Majorana,
  Maksimovic, Malik, Man, Mandic, Mangano, Mansell, Manske, Mantovani,
  Marchesoni, Marion, M{\ifmmode\acute{a}\else\'{a}\fi}rka,
  M{\ifmmode\acute{a}\else\'{a}\fi}rka, Markakis, Markosyan, Markowitz, Maros,
  Marquina, Marsat, Martelli, Martin, Martin, Martynov, Mason, Massera,
  Masserot, Massinger, Masso-Reid, Mastrogiovanni, Matas, Matichard, Matone,
  Mavalvala, Mazumder, McCann, McCarthy, McClelland, McCormick, McCuller,
  McGuire, McIver, McManus, McRae, McWilliams, Meacher, Meadors, Mehmet, Mehta,
  Meidam, Melatos, Mendell, Mercer, Mereni, Merilh, Merzougui, Meshkov,
  Messenger, Messick, Metzdorff, Meyers, Miao, Michel, Middleton, Mikhailov,
  Milano, Miller, Miller, Millhouse, Mills, Milovich-Goff, Minazzoli, Minenkov,
  Mishkin, Mishra, Mistry, Mitra, Mitrofanov, Mitselmakher, Mittleman, Mo,
  Moffa, Mogushi, Mohapatra, Montani, Moore, Moraru, Moreno, Morisaki, Mours,
  Mow-Lowry, Mukherjee, Mukherjee, Mukherjee, Mukund, Mullavey, Munch,
  Mu{\ifmmode\tilde{n}\else\~{n}\fi}iz, Muratore, Murray, Nagar, Nardecchia,
  Naticchioni, Nayak, Neilson, Nelemans, Nelson, Nery, Neunzert, Ng, Ng,
  Nguyen, Nichols, Nissanke, Nocera, North, Nuttall, Obergaulinger, Oberling,
  O{'}Brien, O{'}Dea, Ogin, Oh, Oh, Ohme, Ohta, Okada, Oliver, Oppermann, Oram,
  O{'}Reilly, Ormiston, Ortega, O{'}Shaughnessy, Ossokine, Ottaway, Overmier,
  Owen, Pace, Pagano, Page, Pai, Pai, Palamos, Palashov, Palomba, Pal-Singh,
  Pan, Pang, Pang, Pankow, Pannarale, Pant, Paoletti, Paoli, Parida, Parker,
  Pascucci, Pasqualetti, Passaquieti, Passuello, Patil, Patricelli, Pearlstone,
  Pedersen, Pedraza, Pedurand, Pele, Penn, Perez, Perreca, Pfeiffer, Phelps,
  Phukon, Piccinni, Pichot, Piergiovanni, Pillant, Pinard, Pirello, Pitkin,
  Poggiani, Pong, Ponrathnam, Popolizio, Porter, Powell, Prajapati, Prasad,
  Prasai, Prasanna, Pratten, Prestegard, Privitera, Prodi, Prokhorov, Puncken,
  Punturo, Puppo, P{\ifmmode\ddot{u}\else\"{u}\fi}rrer, Qi, Quetschke,
  Quinonez, Quintero, Quitzow-James, Raab, Radkins, Radulescu, Raffai, Raja,
  Rajan, Rajbhandari, Rakhmanov, Ramirez, Ramos-Buades, Rana, Rao, Rapagnani,
  Raymond, Razzano, Read, Regimbau, Rei, Reid, Reitze, Ren, Ricci, Richardson,
  Richardson, Ricker, Riles, Rizzo, Robertson, Robie, Robinet, Rocchi, Rolland,
  Rollins, Roma, Romanelli, Romano, Romel, Romie, Rose,
  Rosi{\ifmmode\acute{n}\else\'{n}\fi}ska, Rosofsky, Ross, Rowan,
  R{\ifmmode\ddot{u}\else\"{u}\fi}diger, Ruggi, Rutins, Ryan, Sachdev, Sadecki,
  Sakellariadou, Salconi, Saleem, Samajdar, Sammut, Sanchez, Sanchez,
  Sanchis-Gual, Sandberg, Sanders, Santiago, Sarin, Sassolas, Sathyaprakash,
  Saulson, Sauter, Savage, Schale, Scheel, Scheuer, Schmidt, Schnabel,
  Schofield, Sch{\ifmmode\ddot{o}\else\"{o}\fi}nbeck, Schreiber, Schulte,
  Schutz, Schwalbe, Scott, Scott, Seidel, Sellers, Sengupta, Sennett, Sentenac,
  Sequino, Sergeev, Setyawati, Shaddock, Shaffer, Shahriar, Shaner, Shao,
  Sharma, Shawhan, Shen, Shink, Shoemaker, Shoemaker, ShyamSundar, Siellez,
  Sieniawska, Sigg, Silva, Singer, Singh, Singhal, Sintes, Sitmukhambetov,
  Skliris, Slagmolen, Slaven-Blair, Smith, Smith, Somala, Son, Sorazu,
  Sorrentino, Souradeep, Sowell, Spencer, Srivastava, Srivastava, Staats,
  Stachie, Standke, Steer, Steinke, Steinlechner, Steinlechner, Steinmeyer,
  Stevenson, Stocks, Stone, Stops, Strain, Stratta, Strigin, Strunk, Sturani,
  Stuver, Sudhir, Summerscales, Sun, Sunil, Suresh, Sutton, Swinkels,
  Szczepa{\ifmmode\acute{n}\else\'{n}\fi}czyk, Tacca, Tait, Talbot, Talukder,
  Tanner, T{\ifmmode\acute{a}\else\'{a}\fi}pai, Taracchini, Tasson, Taylor,
  Thies, Thomas, Thomas, Thondapu, Thorne, Thrane, Tiwari, Tiwari, Tiwari,
  Toland, Tonelli, Tornasi, Torres-Forn{\ifmmode\acute{e}\else\'{e}\fi},
  Torrie, T{\ifmmode\ddot{o}\else\"{o}\fi}yr{\ifmmode\ddot{a}\else\"{a}\fi},
  Travasso, Traylor, Tringali, Trovato, Trozzo, Trudeau, Tsang, Tse, Tso,
  Tsukada, Tsuna, Tuyenbayev, Ueno, Ugolini, Unnikrishnan, Urban, Usman,
  Vahlbruch, Vajente, Valdes, van Bakel, van Beuzekom, van~den Brand, Van
  Den~Broeck, Vander-Hyde, van Heijningen, van~der Schaaf, van Veggel, Vardaro,
  Varma, Vass, Vas{\ifmmode\acute{u}\else\'{u}\fi}th, Vecchio, Vedovato,
  Veitch, Veitch, Venkateswara, Venugopalan, Verkindt, Vetrano,
  Vicer{\ifmmode\acute{e}\else\'{e}\fi}, Viets, Vine, Vinet, Vitale, Vo, Vocca,
  Vorvick, Vyatchanin, Wade, Wade, Wade, Walet, Walker, Wallace, Walsh, Wang,
  Wang, Wang, Wang, Wang, Ward, Warden, Warner, Was, Watchi, Weaver, Wei,
  Weinert, Weinstein, Weiss, Wellmann, Wen, Wessel, We{\ss}els, Westhouse,
  Wette, Whelan, Whiting, Whittle, Wilken, Williams, Williamson, Willis,
  Willke, Wimmer, Winkler, Wipf, Wittel, Woan, Woehler, Wofford, Worden,
  Wright, Wu, Wysocki, Xiao, Yamamoto, Yancey, Yang, Yap, Yazback, Yeeles, Yu,
  Yu, Yuen, Yvert, Zadro{\ifmmode\dot{z}\else\.{z}\fi}ny, Zanolin, Zelenova,
  Zendri, Zevin, Zhang, Zhang, Zhang, Zhao, Zhou, Zhou, Zhu, Zucker, Zweizig,
  and {The LIGO Scientific Collaboration and the Virgo
  Collaboration}]{Abbott2020Feb}
Abbott,~B.~P. \latin{et~al.}  {A guide to LIGO{\textendash}Virgo detector noise
  and extraction of transient gravitational-wave signals}. \emph{Classical
  Quantum Gravity} \textbf{2020}, \emph{37}, 055002\relax
\mciteBstWouldAddEndPuncttrue
\mciteSetBstMidEndSepPunct{\mcitedefaultmidpunct}
{\mcitedefaultendpunct}{\mcitedefaultseppunct}\relax
\EndOfBibitem
\bibitem[Bradbury \latin{et~al.}(2018)Bradbury, Frostig, Hawkins, Johnson,
  Leary, Maclaurin, Necula, Paszke, Vander{P}las, Wanderman-{M}ilne, and
  Zhang]{Jax}
Bradbury,~J.; Frostig,~R.; Hawkins,~P.; Johnson,~M.~J.; Leary,~C.;
  Maclaurin,~D.; Necula,~G.; Paszke,~A.; Vander{P}las,~J.;
  Wanderman-{M}ilne,~S.; Zhang,~Q. {JAX}: composable transformations of
  {P}ython+{N}um{P}y programs. 2018; \url{http://github.com/google/jax}\relax
\mciteBstWouldAddEndPuncttrue
\mciteSetBstMidEndSepPunct{\mcitedefaultmidpunct}
{\mcitedefaultendpunct}{\mcitedefaultseppunct}\relax
\EndOfBibitem
\bibitem[Maclaurin \latin{et~al.}(2015)Maclaurin, Duvenaud, and
  Adams]{Autograd}
Maclaurin,~D.; Duvenaud,~D.; Adams,~R.~P. Autograd: Effortless gradients in
  numpy. ICML 2015 AutoML Workshop. 2015; p~5\relax
\mciteBstWouldAddEndPuncttrue
\mciteSetBstMidEndSepPunct{\mcitedefaultmidpunct}
{\mcitedefaultendpunct}{\mcitedefaultseppunct}\relax
\EndOfBibitem
\bibitem[Pilgrim(2021)]{pilgrim_piecewise-regression_2021}
Pilgrim,~C. piecewise-regression (aka segmented regression) in {Python}.
  \emph{J. Open Source Softw.} \textbf{2021}, \emph{6}, 3859\relax
\mciteBstWouldAddEndPuncttrue
\mciteSetBstMidEndSepPunct{\mcitedefaultmidpunct}
{\mcitedefaultendpunct}{\mcitedefaultseppunct}\relax
\EndOfBibitem
\bibitem[Peng(2011)]{peng_reproducible_2011}
Peng,~R.~D. Reproducible {Research} in {Computational} {Science}.
  \emph{Science} \textbf{2011}, \emph{334}, 1226--1227\relax
\mciteBstWouldAddEndPuncttrue
\mciteSetBstMidEndSepPunct{\mcitedefaultmidpunct}
{\mcitedefaultendpunct}{\mcitedefaultseppunct}\relax
\EndOfBibitem
\bibitem[Pilgrim \latin{et~al.}(2023)Pilgrim, Kent, Hosseini, and
  Chalstrey]{Pilgrim2023Apr}
Pilgrim,~C.; Kent,~P.; Hosseini,~K.; Chalstrey,~E. {Ten simple rules for
  working with other people{'}s code}. \emph{PLoS Comput. Biol.} \textbf{2023},
  \emph{19}, e1011031\relax
\mciteBstWouldAddEndPuncttrue
\mciteSetBstMidEndSepPunct{\mcitedefaultmidpunct}
{\mcitedefaultendpunct}{\mcitedefaultseppunct}\relax
\EndOfBibitem
\bibitem[rea(2023)]{readthedocs}
{Read the Docs: documentation simplified}. 2023;
  \url{https://docs.readthedocs.io/en/stable}, [Online; accessed 19. Oct.
  2023]\relax
\mciteBstWouldAddEndPuncttrue
\mciteSetBstMidEndSepPunct{\mcitedefaultmidpunct}
{\mcitedefaultendpunct}{\mcitedefaultseppunct}\relax
\EndOfBibitem
\bibitem[dox(2023)]{doxygen}
{Doxygen: Doxygen}. 2023; \url{https://www.doxygen.nl/index.html}, [Online;
  accessed 19. Oct. 2023]\relax
\mciteBstWouldAddEndPuncttrue
\mciteSetBstMidEndSepPunct{\mcitedefaultmidpunct}
{\mcitedefaultendpunct}{\mcitedefaultseppunct}\relax
\EndOfBibitem
\bibitem[Chacon and Straub(2014)Chacon, and Straub]{git}
Chacon,~S.; Straub,~B. \emph{Pro git}; Apress, 2014\relax
\mciteBstWouldAddEndPuncttrue
\mciteSetBstMidEndSepPunct{\mcitedefaultmidpunct}
{\mcitedefaultendpunct}{\mcitedefaultseppunct}\relax
\EndOfBibitem
\bibitem[git(2023)]{gitlab}
{The DevSecOps Platform}. 2023; \url{https://about.gitlab.com}, [Online;
  accessed 19. Oct. 2023]\relax
\mciteBstWouldAddEndPuncttrue
\mciteSetBstMidEndSepPunct{\mcitedefaultmidpunct}
{\mcitedefaultendpunct}{\mcitedefaultseppunct}\relax
\EndOfBibitem
\bibitem[git(2023)]{github}
{Build software better, together}. 2023; \url{https://github.com}, [Online;
  accessed 19. Oct. 2023]\relax
\mciteBstWouldAddEndPuncttrue
\mciteSetBstMidEndSepPunct{\mcitedefaultmidpunct}
{\mcitedefaultendpunct}{\mcitedefaultseppunct}\relax
\EndOfBibitem
\bibitem[Abbasi(2023)]{SilicoStudio}
Abbasi,~D. {Cutting-Edge Free Tools To Unlock the Power of Computational
  Chemistry - Silico Studio}. \emph{Silico Studio} \textbf{2023}, \relax
\mciteBstWouldAddEndPunctfalse
\mciteSetBstMidEndSepPunct{\mcitedefaultmidpunct}
{}{\mcitedefaultseppunct}\relax
\EndOfBibitem
\bibitem[Mol(2023)]{Molssi}
{MolSSI{'}s Best Practices {\textendash} MolSSI}. 2023;
  \url{https://molssi.org/molssis-best-practices}, [Online; accessed 19. Oct.
  2023]\relax
\mciteBstWouldAddEndPuncttrue
\mciteSetBstMidEndSepPunct{\mcitedefaultmidpunct}
{\mcitedefaultendpunct}{\mcitedefaultseppunct}\relax
\EndOfBibitem
\bibitem[Mis(2023)]{Missingsemester}
{The Missing Semester of Your CS Education}. 2023;
  \url{https://missing.csail.mit.edu}, [Online; accessed 19. Oct. 2023]\relax
\mciteBstWouldAddEndPuncttrue
\mciteSetBstMidEndSepPunct{\mcitedefaultmidpunct}
{\mcitedefaultendpunct}{\mcitedefaultseppunct}\relax
\EndOfBibitem
\bibitem[Learning(2023)]{DeepLearningMIT}
Learning,~M.~D. {MIT Deep Learning 6.S191}. 2023;
  \url{http://introtodeeplearning.com}, [Online; accessed 19. Oct. 2023]\relax
\mciteBstWouldAddEndPuncttrue
\mciteSetBstMidEndSepPunct{\mcitedefaultmidpunct}
{\mcitedefaultendpunct}{\mcitedefaultseppunct}\relax
\EndOfBibitem
\bibitem[Chem(2017)]{tmp}
Chem,~T. {Computational Chemistry 0.1 - Introduction}. 2017;
  \url{https://www.youtube.com/watch?v=YF-amZgE2h4&list=PLm8ZSArAXicIWTHEWgHG5mDr8YbrdcN1K},
  [Online; accessed 19. Oct. 2023]\relax
\mciteBstWouldAddEndPuncttrue
\mciteSetBstMidEndSepPunct{\mcitedefaultmidpunct}
{\mcitedefaultendpunct}{\mcitedefaultseppunct}\relax
\EndOfBibitem
\bibitem[{}(2023)]{vsl}
{}, {Virtual Simulation Lab}. 2023;
  \url{https://www.youtube.com/@VirtualSimulationLab/videos}, [Online; accessed
  19. Oct. 2023]\relax
\mciteBstWouldAddEndPuncttrue
\mciteSetBstMidEndSepPunct{\mcitedefaultmidpunct}
{\mcitedefaultendpunct}{\mcitedefaultseppunct}\relax
\EndOfBibitem
\bibitem[{}(2023)]{statquest}
{}, {StatQuest with Josh Starmer}. 2023;
  \url{https://www.youtube.com/@statquest}, [Online; accessed 19. Oct.
  2023]\relax
\mciteBstWouldAddEndPuncttrue
\mciteSetBstMidEndSepPunct{\mcitedefaultmidpunct}
{\mcitedefaultendpunct}{\mcitedefaultseppunct}\relax
\EndOfBibitem
\bibitem[{}(2023)]{comptoolkit}
{}, {The Computational Toolkit}. 2023;
  \url{https://www.youtube.com/@thecomputationaltoolkit2890/videos}, [Online;
  accessed 19. Oct. 2023]\relax
\mciteBstWouldAddEndPuncttrue
\mciteSetBstMidEndSepPunct{\mcitedefaultmidpunct}
{\mcitedefaultendpunct}{\mcitedefaultseppunct}\relax
\EndOfBibitem
\bibitem[Cumby \latin{et~al.}(2023)Cumby, Degiacomi, Erastova, Güven, Hobday,
  Mey, Pollak, and Szabla]{CumbyJOSE2023}
Cumby,~J.; Degiacomi,~M.; Erastova,~V.; Güven,~J.; Hobday,~C.; Mey,~A.;
  Pollak,~H.; Szabla,~R. Course Materials for an Introduction to Data-Driven
  Chemistry. \emph{J. Open Source Educ.} \textbf{2023}, \emph{6}, 192\relax
\mciteBstWouldAddEndPuncttrue
\mciteSetBstMidEndSepPunct{\mcitedefaultmidpunct}
{\mcitedefaultendpunct}{\mcitedefaultseppunct}\relax
\EndOfBibitem
\bibitem[French(2023)]{FrenchJOSE2023}
French,~T. R for Data Analysis: An open-source resource for teaching and
  learning analytics with R. \emph{J. Open Source Educ.} \textbf{2023},
  \emph{6}, 202\relax
\mciteBstWouldAddEndPuncttrue
\mciteSetBstMidEndSepPunct{\mcitedefaultmidpunct}
{\mcitedefaultendpunct}{\mcitedefaultseppunct}\relax
\EndOfBibitem
\bibitem[Storopoli \latin{et~al.}(2021)Storopoli, Huijzer, and
  Alonso]{JuliaDataScience}
Storopoli,~J.; Huijzer,~R.; Alonso,~L. \emph{Julia Data Science}; 2021;
  \url{https://juliadatascience.io}\relax
\mciteBstWouldAddEndPuncttrue
\mciteSetBstMidEndSepPunct{\mcitedefaultmidpunct}
{\mcitedefaultendpunct}{\mcitedefaultseppunct}\relax
\EndOfBibitem
\bibitem[White(2021)]{white2021deep}
White,~A.~D. Deep Learning for Molecules and Materials. \emph{Living Journal of
  Computational Molecular Science} \textbf{2021}, \emph{3}, 1499\relax
\mciteBstWouldAddEndPuncttrue
\mciteSetBstMidEndSepPunct{\mcitedefaultmidpunct}
{\mcitedefaultendpunct}{\mcitedefaultseppunct}\relax
\EndOfBibitem
\end{mcitethebibliography}

\providecommand{\latin}[1]{#1}
\makeatletter
\providecommand{\doi}
  {\begingroup\let\do\@makeother\dospecials
  \catcode`\{=1 \catcode`\}=2 \doi@aux}
\providecommand{\doi@aux}[1]{\endgroup\texttt{#1}}
\makeatother
\providecommand*\mcitethebibliography{\thebibliography}
\csname @ifundefined\endcsname{endmcitethebibliography}
  {\let\endmcitethebibliography\endthebibliography}{}

\end{document}